\documentclass[aps,prd,preprint,superscriptaddress,nofootinbib]{revtex4-1}
\usepackage{bm}
\usepackage{indentfirst}
\usepackage{amsmath}
\usepackage{graphicx}
\usepackage{amssymb}
\usepackage{subfigure}
\usepackage{amssymb}
\usepackage{hyperref}
\hypersetup{
  colorlinks=true,
  linkcolor=red,
  citecolor=blue,
}
\usepackage[utf8]{inputenc}
\begin{document}
\title{Gauge transformation of scalar induced gravitational waves}
\author{Yizhou Lu}
\email{louischou@hust.edu.cn}
\affiliation{School of Physics, Huazhong University of Science and Technology,
Wuhan, Hubei 430074, China}
\author{Arshad Ali}
\email{aa$_$math@hust.edu.cn}
\affiliation{School of Physics, Huazhong University of Science and Technology,
Wuhan, Hubei 430074, China}
\author{Yungui Gong}
\email{Corresponding author. yggong@hust.edu.cn}
\affiliation{School of Physics, Huazhong University of Science and Technology,
Wuhan, Hubei 430074, China}
\author{Jiong Lin}
\email{jionglin@hust.edu.cn}
\affiliation{School of Physics, Huazhong University of Science and Technology,
Wuhan, Hubei 430074, China}
\author{Fengge Zhang}
\email{fenggezhang@hust.edu.cn}
\affiliation{School of Physics, Huazhong University of Science and Technology,
Wuhan, Hubei 430074, China}

\begin{abstract}
The gauge dependence of the scalar induced gravitational
waves (SIGWs) generated at the second order
imposes a challenge to the discussion of the secondary gravitational waves
generated by scalar perturbations.
We provide a general formula that is valid in any gauge for the calculation
of SIGWs and the relationship for SIGWs calculated in various
gauges under the coordinate transformation.
The formula relating SIGWs in the Newtonian gauge
to other gauges is used to calculate SIGWs in six different gauges.
We find that the Newtonian gauge, the uniform curvature gauge,
the synchronous gauge and the uniform expansion gauge yield
the same result for the energy density of SIGWs.
We also identify and eliminate the pure gauge modes that exist in the
synchronous gauge. 
In the total matter gauge and the comoving orthogonal gauge,
the energy density of SIGWs increases as $\eta^2$. While in the uniform density gauge,
the energy density of SIGWs increases as $\eta^6$.
\end{abstract}

\preprint{2006.03450}

\maketitle

\section{Introduction}

The detection of gravitational waves (GWs) by the LIGO Collaboration
and the Virgo Collaboration opens up a new avenue for probing the property
of gravity in the strong field and nonlinear regions \cite{Abbott:2016blz,Abbott:2016nmj,
	Abbott:2017vtc,Abbott:2017oio,TheLIGOScientific:2017qsa,
	Abbott:2017gyy,LIGOScientific:2018mvr}.
Although the primordial GWs may be too small to be detected
even by the third generation ground based and the spaced
based GW detectors, the scalar induced GWs (SIGWs) generated
at the second order can be large enough because the
amplitude of the primordial scalar
power spectrum is constrained by the cosmic microwave background anisotropy measurements to be $\sim 10^{-9}$ only at large scales
and it can be as large as $\sim 0.01$ at small scales \cite{Akrami:2018odb}.
The SIGWs \cite{Ananda:2006af,Baumann:2007zm,Bugaev:2009zh,Bugaev:2010bb,Inomata:2016rbd,
Gong:2017qlj,Cai:2018dig,Kohri:2018awv,Lu:2019sti,Cai:2019amo,Cai:2019elf,Drees:2019xpp,
Inomata:2018epa,Inomata:2019ivs,Inomata:2019zqy,Espinosa:2018eve,Orlofsky:2016vbd,
Garcia-Bellido:2017aan,Garcia-Bellido:2017mdw,
Cheng:2018yyr,Cai:2019jah,Yuan:2019wwo,
DeLuca:2019ufz,Fu:2019vqc,Hajkarim:2019nbx,Cai:2019cdl,Domenech:2019quo,Lin:2020goi,Domenech:2020kqm,Braglia:2020eai} produced by the large scalar perturbations during
the radiation dominated era can be much larger than the primordial GWs
and they have peak frequency at nanohertz or millihertz which
can be detected by the space based GW observatory like Laser Interferometer Space Antenna (LISA) \cite{Danzmann:1997hm,Audley:2017drz}, TianQin \cite{Luo:2015ght}
and TaiJi \cite{Hu:2017mde}, and the Pulsar Timing Array (PTA) \cite{Kramer:2013kea,Hobbs:2009yy,McLaughlin:2013ira,Hobbs:2013aka}
including the Square Kilometer Array (SKA) \cite{Moore:2014lga} in the future.
In the literature, SIGWs are also called secondary GWs. This is because,
in addition to the first-order scalar perturbation as the source
of second-order tensor perturbation, there are other
sources coming from the first-order vector and tensor perturbations
such as the scalar-vector, scalar-tensor, vector-vector,
vector-tensor, and tensor-tensor combinations \cite{Gong:2019mui}.
In order to distinguish GWs produced by these sources, we use SIGWs in this paper.

Unlike the primordial GWs which are the first-order tensor perturbations
and gauge invariant, SIGWs sourced by the first-order scalar perturbations due to the nonlinearity of Einstein's equation are gauge dependent \cite{Matarrese:1997ay,Bruni:1996im,Malik:2008im,Hwang:2017oxa,Domenech:2017ems,Yuan:2019fwv,Tomikawa:2019tvi}.
Apart from the issue of the choice of physical gauge 
and the true observable measured by GW detectors,
the gauge dependence requires us to calculate SIGWs
in each gauge. To avoid this problem, 
in this paper we discuss the relationship 
for SIGWs in various gauges under the coordinate transformation.
Although it was discussed in the literature, the formula
was derived without including the contribution of the first-order
perturbation $E$ \cite{Gong:2019mui,Hwang:2017oxa,DeLuca:2019ufz}. 
For this reason, the results of SIGWs obtained in the synchronous gauge cannot
be derived from other gauges by the coordinate transformation
although their results on the energy density agree with each other \cite{Yuan:2019fwv,Inomata:2019yww,DeLuca:2019ufz}.
We provide a general formula that is valid in any gauge 
for the calculation of SIGWs
and the relationship for SIGWs calculated in various gauges under the coordinate transformation.

The paper is organized as follows. 
In the next section, we review the basic formulas to calculate SIGWs 
and discuss the gauge transformation. 
We use the \emph{mathematica} package \emph{xPand} \cite{xact,Pitrou:2013hga}
to derive some equations.
We also provide the prescription to obtain the expressions 
in other gauges from the result in the Newtonian gauge by 
using the gauge transformation of the second-order tensor perturbation.
In Sec. \ref{sec.3}, we use the prescription presented in
Sec. \ref{sec.2} to derive the kernels in
the uniform curvature gauge,
the synchronous gauge, the uniform expansion gauge,
the total matter gauge, the comoving orthogonal gauge
and the uniform density gauge.
The late time behaviors of SIGWs in these gauges are then analyzed.
We also calculate the kernels directly in the uniform curvature gauge, 
the synchronous gauge and the total matter gauge, and show that
they agree with those obtained from the coordinate transformations.
The pure gauge modes in the synchronous gauge are identified.
We conclude the paper in Sec. \ref{sec4}.
The discussion on the perturbations is presented in the Appendix. 

\section{Basics of SIGWs and gauge transformations}
\label{sec.2}

We consider the following perturbed metric
\begin{equation}
ds^{2}=a^2\left[-(1+2 \phi) \mathrm{d} \eta^{2}+2B_{,i} \mathrm{d} x^{i} \mathrm{d} \eta+\left((1-2\psi)\delta_{i j}+2E_{,ij}+\frac12 h_{ij}^{\mathrm{TT}}\right) \mathrm{d} x^{i} \mathrm{d} x^{j}\right],
\end{equation}
where the scalar perturbations $\phi$, $\psi$, $B$ and $E$ are first order,
but the transverse traceless (TT) part $h_{ij}^{\mathrm{TT}}$ is second order for the purpose of calculating the scalar induced tensor perturbations, $h^{\mathrm{TT}}_{ii}=0$ and $\partial_i h^\mathrm{TT}_{ij}=0$. The first-order vector and tensor perturbations are not considered.

Perturbing Einstein's equation $G_{\mu\nu}=8\pi G T_{\mu\nu}$ to the second order, we get the contribution of the first-order scalar perturbations to the second-order tensor perturbation as
\begin{align}
\label{theeq}
  h_{ij}^{\mathrm{TT}\prime\prime}+2\mathcal{H}h_{ij}^{\mathrm{TT}\prime}-\nabla^2h_{ij}^{\mathrm{TT}}=4\mathcal{T}_{ij}^{lm}s_{lm},
\end{align}
where $\mathcal{H}=a^\prime/a$, the prime denotes the derivative with respect to the conformal time $\eta$,
and the projection tensor $\mathcal{T}_{ij}^{lm}$ extracting the transverse, trace-free part of a tensor will be discussed explicitly below. The source $s_{ij}$ is
\begin{equation}
\label{source}
\begin{split}
 -s_{ij}=&\psi_{,i}\psi_{,j}+\phi_{,i}\phi_{,j}
 -\sigma_{,ij}\left(\phi^\prime+\psi^\prime-\nabla^2\sigma\right)
 +\left(\psi_{,i}^\prime\sigma_{,j}+\psi_{,j}^\prime\sigma_{,i}\right)
 -\sigma_{,ik}\sigma_{,jk}+2\psi_{,ij}\left(\phi+\psi\right)\\
 &-8\pi Ga^2({\rho_0}+{P_0})\delta V_{,i}\delta V_{,j}-2\psi_{,ij}\nabla^2E
 +2E_{,ij}\left(\psi^{\prime\prime}+2\mathcal{H}\psi^\prime-\nabla^2\psi\right)-E_{,ik}^\prime E_{,jk}^\prime\\
 &+E_{,ikl}E_{,jkl}+2\left(\psi_{,jk}E_{,ik}+\psi_{,ik}E_{,jk}\right)
 -2\mathcal{H}(\psi_{,i}E_{,j}^\prime+\psi_{,j}E_{,i}^\prime)-\left(\psi_{,i}^\prime E_{,j}^\prime+\psi_{,j}^\prime E_{,i}^\prime\right)\\
 &
 -\left(\psi_{,i}E_{,j}^{\prime\prime}+\psi_{,j}E_{,i}^{\prime\prime}\right)
 +2E_{,ij}^\prime\psi^\prime+E_{,ijk}\left(E^{\prime\prime}+2\mathcal{H}E^\prime-
 \nabla^2E\right)_{,k},
 \end{split}
 \end{equation}
where $\sigma= E^\prime-B$ is the shear potential,
the anisotropic stress tensor $\Pi_{ij}$ is assumed to be zero,
$\delta V$ is the scalar part of the velocity perturbation
of the fluid, and
$\rho_0$ and $P_0$ are the background values
of energy density and pressure of the fluid.
The detailed discussion of these variables is presented in the Appendix.
In gauges with $E=0$, the above equation \eqref{source}
reduces to the results given in \cite{Gong:2019mui,DeLuca:2019ufz,Hwang:2017oxa} with vanishing anisotropic stress. In general,
we need to use Eq. \eqref{source} instead. In particular, we should include all the terms involving $E$ in the synchronous gauge.

For GWs propagating along the direction $\bm{k}$, we introduce
the orthonormal bases ${\mathbf e}$ and $\bar{\mathbf e}$ with
${\bm k}\cdot {\mathbf e}={\bm k}\cdot \bar{\mathbf e}={\mathbf e}\cdot \bar{\mathbf e}=0$ and $|{\mathbf e}|=|\bar{\mathbf e}|=1$,
then the plus and cross polarization tensors are expressed as
 \begin{equation}
 \label{poltensor1}
 \begin{split}
      \mathbf e^+_{ij}=&\frac{1}{\sqrt{2}}(\mathbf e_i \mathbf e_j-\mathbf e_i \mathbf e_j),\\
      \mathbf e_{ij}^\times=&\frac{1}{\sqrt{2}}(\mathbf e_i\bar{\mathbf e}_j+\bar{\mathbf e}_i \mathbf e_j).
 \end{split}
 \end{equation}\textit{\(\)}
The polarization tensors \eqref{poltensor1} are transverse and traceless because $k_i \mathbf e^+_{ij}=k_i \mathbf e^\times_{ij}=0$ and
$\mathbf e^+_{ii}=\mathbf e^\times_{ii}=0$, and they can be used to expand $h^{\mathrm{TT}}_{ij}$,
 \begin{equation}
 \label{hijkeq1}
      h_{ij}^{\mathrm{TT}}(\bm x,\eta)=\int\frac{\mathrm{d}^3k}{(2\pi)^{3/2}}e^{i{\bm k}\cdot {\bm x}}[h^+_{\bm k}(\eta)\mathbf e^+_{ij}+{h}^\times_{\bm k}(\eta) \mathbf e^\times_{ij}].
 \end{equation}
The projection tensor is
 \begin{equation}
 \label{projeq}
     \mathcal{T}_{ij}^{lm}s_{lm}=\int\frac{\mathrm{d}^3 k}{(2\pi)^{3/2}}e^{i {\bm k} \cdot {\bm x}}[\mathbf e_{ij}^+ \mathbf e^{+lm}+\mathbf e_{ij}^\times \mathbf e^{\times lm}]s_{lm}(\bm k,\eta),
 \end{equation}
 where $s_{ij}(\bm k,\eta)$ is the Fourier transformation of $s_{ij}(\bm x,\eta)$.

Assuming equal contributions from the two polarizations, we can use one polarization to calculate its energy density and obtain the total energy density by doubling it.
Working in Fourier space,
the solution to Eq. \eqref{theeq} for the plus polarization $e_{ij}^+$ is
\begin{equation}
\label{hsolution}
  h^+_{\bm k}(\eta)=4 \int\frac{\mathrm{d}^3p}{(2\pi)^{3/2}} \mathbf e^{+ij}p_ip_j\zeta(\bm p)\zeta(\bm k-\bm p)\frac{1}{k^2}I(u,v,x),
\end{equation}
where $x=k\eta$, $u=p/k$, $v=|\bm k-\bm{p}|/k$, $\zeta(\bm k)=\psi+\mathcal{H}\delta\rho/\rho_0^\prime$ is the primordial curvature perturbation,
$I(u,v,x)$ is given by \cite{Kohri:2018awv,Inomata:2016rbd,Espinosa:2018eve,Lu:2019sti}
\begin{align}
\label{I_int}
    I(u,v,x)=\int_0^x\mathrm{d}\tilde{x}\frac{a(\tilde{\eta})}{a(\eta)}kG_{k}(\eta,\tilde{\eta})f(u,v,\tilde x),
\end{align}
the Green's function $G_{k}(\tilde{\eta},\eta)$ to Eq. \eqref{theeq} is
\begin{align}
G_k(\tilde{\eta},\eta)=\frac{\sin(x-\tilde{x})}{k},
\end{align}
$f(u,v,x)$ is symmetric about $u$ and $v$ and it is related with the source $S_{\bm k}^+=\mathbf{e}^{+ij}s_{ij}(\bm k,\eta)$ as
\begin{equation}
\label{sfrel1}
S_{\bm k}^+(\eta)=\int\frac{\mathrm{d}^3 p}{(2\pi)^{3/2}}\zeta(\bm p)\zeta(\bm k-\bm p)\mathbf{e}^{+ij}p_i p_j f(u,v,x).
\end{equation}
In the above derivation, we assume that the production of induced GWs begins long before the horizon reentry during the radiation domination. In this paper,
we consider the production of SIGWs in the radiation dominated era only.
During radiation domination, $\mathcal{H}\sim\eta^{-1}$,
the power spectrum of SIGWs is given by
\begin{align}
\label{PStensor}
    \mathcal{P}_h(k,x)=4\int_{0}^\infty\mathrm{d}u\int_{|1-u|}^{1+u}\mathrm{d}v
    \left[\frac{4u^2-(1+u^2-v^2)}{4uv}\right]^2I^{2}(u,v,x)\mathcal{P}_\zeta(uk)\mathcal{P}_\zeta(vk)
,
\end{align}
and the fractional energy density of SIGWs is
\begin{align}
    \Omega_{\mathrm{GW}}=\frac{1}{24}\left(
    \frac{k}{\mathcal{H}}
    \right)^2\overline{\mathcal{P}_h(k,x)},
\end{align}
where $\mathcal{P}_\zeta$ is the primordial scalar power spectrum and
$\mathcal{P}_h$ is given by
\begin{align}
\left\langle h_{\bm k_1}^{s_1}(\eta)h_{\bm k_2}^{s_2}(\eta)\right\rangle=\frac{2\pi^2}{k_1^3}\delta_{s_1s_2}\delta^3(\bm k_1+\bm k_2)\mathcal{P}_h(k_1,\eta),\quad s_i=+,\times.
\end{align}

To separate the time evolution, we introduce the transfer function $T$ by defining $\phi(\bm k,\eta)=\phi(\bm k,0)T(\eta)$.
In the Newtonian gauge (also referred as Poisson gauge and zero-shear gauge), $B=E=0$, 
we have $\phi_\mathrm{N}=\psi_\mathrm{N}=\Phi=\Psi$ if the anisotropic stress vanishes,
here $\Phi$ and $\Psi$ are the Bardeen's potentials defined in \eqref{bardeenphi} and \eqref{bardeenpsi},
and during radiation domination $\Phi=\Psi=2\zeta/3$ on superhorizon scales.
Therefore, in the Newtonian gauge, $\phi_\text{N}(\bm k,0)=2\zeta(\bm k)/3$,
the transfer function $T_\mathrm{N}(x)$ with the initial condition $T_\mathrm{N}(0)=1$  is
\begin{align}\label{T_N}
    T_{\mathrm{N}}(x)=\frac{9}{x^2}\left(
    \frac{\sin(x/\sqrt{3})}{x/\sqrt{3}}-\cos(x/\sqrt{3})
    \right),
\end{align}
and we have
\begin{gather}\label{fNewton}
    f_{\mathrm{N}}(u,v,x)=2T_\mathrm{N}(vx)T_\mathrm{N}(ux)+[T_\mathrm{N}(vx)+vxT_\mathrm{N}^*(vx)][T_\mathrm{N}(ux)+uxT_\mathrm{N}^*(ux)].
\end{gather}
The explicit expression for $I_{\mathrm{N}}(u,v,x)$ is\cite{Kohri:2018awv,Espinosa:2018eve}
\begin{equation}
\label{I_N}
\begin{split}
I_{\mathrm{N}}(u,v,x)=&\frac{3}{4u^3v^3x}\left(
-\frac{4}{x^3}\left(
(u^2+v^2-3)uvx^3\sin x-6uvx^2\cos\frac{ux}{\sqrt{3}}\cos\frac{vx}{\sqrt{3}}\right.\right.\\
&+6\sqrt{3}ux\cos\frac{ux}{\sqrt{3}}\sin\frac{vx}{\sqrt{3}}+6\sqrt{3}vx\sin\frac{ux}{\sqrt{3}}\cos\frac{vx}{\sqrt{3}}\\
&\left.-3(6+(u^2+v^2-3)x^2)\sin\frac{ux}{\sqrt{3}}\sin\frac{vx}{\sqrt{3}}
\right)+(u^2+v^2-3)^2\\
&\times\left[
\left(
\mathrm{Ci}\left[\left(1+\frac{u-v}{\sqrt{3}}\right)x\right]+
\mathrm{Ci}\left[\left(1+\frac{v-u}{\sqrt{3}}\right)x\right]-
\mathrm{Ci}\left[\left(1+\frac{u+v}{\sqrt{3}}\right)x\right]\right.\right.\\
&\left.
-\mathrm{Ci}\left[\left|1-\frac{u+v}{\sqrt{3}}\right|x\right]
+\ln\left[
\left|\frac{3-(u+v)^2}{3-(u-v)^2}\right|
\right]\right)\sin x+\left(
-\mathrm{Si}\left[\left(1+\frac{u-v}{\sqrt{3}}\right)x\right]\right.\\
&\left.\left.\left.-
\mathrm{Si}\left[\left(1+\frac{v-u}{\sqrt{3}}\right)x\right]+
\mathrm{Si}\left[\left(1-\frac{u+v}{\sqrt{3}}\right)x\right]+
\mathrm{Si}\left[\left(1+\frac{u+v}{\sqrt{3}}\right)x\right]
\right)\cos x
\right]
\right).
\end{split}
\end{equation}
The subscript ``$\mathrm{N}$" indicates that they are evaluated in the Newtonian gauge.
The evolution of $I^2_\text{N}(u,v,x)$ with $u=v=1$ is shown in Fig. \ref{kenerluv1}.
Note that $I_{\mathrm{N}}(u,v,x\rightarrow\infty)\propto x^{-1}$,
and $\Omega_{\mathrm{GW}}(k,x\rightarrow\infty)$ is a constant.
This implies that SIGWs behave like free radiation deep within horizon.

Now we discuss the gauge transformation.
The infinitesimal coordinate transformation is
$x^\mu\to x^\mu+\epsilon^\mu$ with $\epsilon^\mu=[\alpha,\delta^{ij}\partial_j\beta]$.
For the discussion of SIGWs, we do not consider the vector degrees of freedom for the coordinate transformation, and the scalars $\alpha$ and $\beta$ are of first order.
Since the gauge transformation of tensor modes does not depend on the coordinate transformation of the same order, we do not need to consider the second-order coordinate transformation. 
For the second-order tensor perturbation, we have \cite{Malik:2008im, Bruni:1996im}
 \begin{equation}
 \label{gaugetranf1}
     h_{ij}^{\mathrm{TT}}\to h_{ij}^{\mathrm{TT}}+\chi_{ij}^{\mathrm{TT}},
 \end{equation}
where
\begin{equation}
\label{xijtt}
    \chi_{ij}^{\mathrm{TT}}(\bm x,\eta)=\mathcal{T}_{ij}^{lm}\chi_{lm}=\int\frac{\mathrm{d}^3 k}{(2\pi)^{3/2}}e^{i\bm k\cdot \bm x}[\chi^+_{\bm k}(\eta) \mathbf e_{ij}^++{\chi}^\times_{\bm k}(\eta)\mathbf e_{ij}^\times],
\end{equation}
 \begin{align}
\label{chi_fourier}
   \chi^+_{\bm k}(\eta)=&-\int\frac{\mathrm{d}^3p}{(2\pi)^{3/2}} \mathbf e^{+ij}p_ip_j\left(4\alpha(\bm p)\sigma(\bm k-\bm p)+8\mathcal{H}\alpha(\bm p)[E(\bm k-\bm p)+\beta(\bm k-\bm p)]\right.\notag\\&\left.
   +\bm{p}\cdot
   (\bm k-\bm p)\beta(\bm p)[4E(\bm k-\bm p)+2\beta(\bm k-\bm p)]-8\psi(\bm p)\beta(\bm k-\bm p)+2\alpha(\bm p)\alpha(\bm k-\bm p)\right),\notag\\
   =&4\int\frac{\mathrm{d}^3p}{(2\pi)^{3/2}}\mathbf e^{+ij}p_ip_j \zeta(\bm p)\zeta(\bm k-\bm p)\frac{1}{k^2}I_\chi(u,v,x),
 \end{align}
 \begin{equation}
 \label{Ichi}
 \begin{split}
   I_\chi(u,v,x)= &- \frac{1}{9uv}\left[\vphantom{\frac{u^2}{v^2}}2T_\alpha(ux)T_\sigma(vx)
   +2T_\alpha(vx)T_\sigma(ux)+2T_\alpha(ux)T_\alpha(vx)\right.\\
   &-4\left(\frac{u}{v}T_\psi(ux)T_\beta(vx)+\frac{v}{u}T_\psi(vx)T_\beta(ux)\right)\\
   &+\frac{1-u^2-v^2}{uv}\left[T_\beta(ux)T_E(vx)+T_\beta(vx)T_E(ux)
   +T_\beta(ux)T_\beta(vx)\right]\\
   &+4\frac{\mathcal{H}}{k}
   \left(\frac1v T_\alpha(ux)T_E(vx)+\frac1u T_E(ux)T_\alpha(vx)\right.\\
   &\left.\left.\quad +\frac1v T_\alpha(ux)T_\beta(vx)+\frac1u T_\beta(ux)T_\alpha(vx)\right)\right].
 \end{split}
 \end{equation}
We have symmetrized $I_\chi(u,v,x)$ under $u\leftrightarrow v$. Note that the first-order scalar coordinate transformation appears in the transformed second-order tensor perturbations.
With the gauge transformation \eqref{gaugetranf1} and the result for SIGWs in the Newtonian gauge, it is straightforward to derive the semianalytic expression for SIGWs in other gauges without performing the detailed calculation in that gauge.

Combining Eqs. \eqref{hijkeq1}, \eqref{hsolution}, \eqref{gaugetranf1}, \eqref{xijtt} and \eqref{chi_fourier}, we get the following gauge transformation
\begin{align}\label{hchi}
  h^+_{\bm k}\rightarrow h^+_{\bm k}+\chi^+_{\bm k}=4\int\frac{\mathrm{d}^3p}{(2\pi)^{3/2}}\mathbf e^{+ij}(\bm k)p_ip_j\zeta(\bm p)\zeta(\bm k-\bm p)\frac{1}{k^2}\left[I(u,v,x)+I_\chi(u,v,x)\right].
\end{align}
and the transfer functions $T(x)$ are
 \begin{gather}
 \label{defs}
      \alpha(\bm k,x)=\frac{2}{3}\zeta(\bm k)\frac{1}{k}T_\alpha(x),\\
   \beta(\bm k,x)=\frac{2}{3}\zeta(\bm k)\frac{1}{k^2}T_\beta(x),\\
   \sigma(\bm k,x)=\frac{2}{3}\zeta(\bm k)\frac{1}{k}T_\sigma(x),\\
   E(\bm k,x)=\frac{2}{3}\zeta(\bm k)\frac{1}{k^2}T_E(x),\\
   B(\bm k,x)=\frac{2}{3}\zeta(\bm k)\frac{1}{k}T_B(x),\\
   \psi(\bm k,x)=\frac{2}{3}\zeta(\bm k)T_\psi(x),\\
   \phi(\bm k,x)=\frac{2}{3}\zeta(\bm k)T_\phi(x).
 \end{gather}
This gauge transformation \eqref{hchi} is the main result of our paper. 
It shows how the solution or the power spectrum of SIGWs transforms
under the gauge transformation. For example, with the solution in the Newtonian gauge \cite{Kohri:2018awv,Espinosa:2018eve},
we can obtain the solution in any other gauge by replacing the
Newtonian gauge kernel $I_\mathrm{N}(u,v,x)$ in Eq. \eqref{PStensor} according to the following rule
\begin{align}\label{IN}
  I_\mathrm{N}(u,v,x) \to I_{\mathrm N}(u,v,x)+I_\chi(u,v,x),
\end{align}
where
\begin{equation}
\label{ichi_newton}
\begin{split}
   I_\chi(u,v,x)= &- \frac{1}{9uv}\left[\vphantom{\frac{u^2}{v^2}}
   -4\left(\frac{u}{v}T_\mathrm{N}(ux)T_\beta(vx)
   +\frac{v}{u}T_\mathrm{N}(vx)T_\beta(ux)\right)
   +2T_\alpha(ux)T_\alpha(vx)\right.\\
   &\left.+4\frac{\mathcal{H}}{k}
   \left(\frac1v T_\alpha(ux)T_\beta(vx)+\frac1u T_\beta(ux)T_\alpha(vx)\right)
+\frac{1-u^2-v^2}{uv}T_\beta(ux)T_\beta(vx)\right],
\end{split}
\end{equation}
which is obtained by substituting the transfer functions $T_\sigma=T_E=0$ and $T_\psi=T_\text{N}$
in the Newtonian gauge into Eq. \eqref{Ichi},
and the coordinate transformations from the Newtonian
gauge to the other gauge give the transfer functions $T_\alpha$ and $T_\beta$.

\section{The kernel in various gauges}
\label{sec.3}

In this section, we derive the analytic expressions for the kernel $I$
in six other gauges by the coordinate transformation from the Newtonian gauge to the other gauges.
To show the effectiveness of the method by the coordinate transformation,
we also calculate the kernels directly in the uniform curvature gauge, 
the synchronous gauge and the comoving gauge (total matter gauge), 
and show that they agree with
the expressions obtained from the coordinate transformations.
In some gauges, the residual gauge transformations are used to eliminate the pure gauge modes.
We then discuss the late time limit of $\Omega_{\text{GW}}$ in those gauges.

\subsection{Uniform curvature gauge}

The uniform curvature gauge is also called the flat gauge. In this gauge, $\psi=E=0$ and the transfer functions are
\begin{align}
T_\phi(x)=&\frac{3\sin(x/\sqrt{3})}{x/\sqrt{3}},\\
T_B(x)=&-\frac{9}{x}\left[
\frac{\sin(x/\sqrt{3})}{x/\sqrt{3}}-\cos(x/\sqrt{3})
\right],
\end{align}
where we assume the same initial condition as that
in the Newtonian gauge for the
gauge-invariant perturbation $\zeta$ that
is conserved well outside the horizon.
Combining Eqs. \eqref{source} and \eqref{sfrel1} , we obtain
\begin{equation}
\label{fuceq1}
\begin{split}
f_{\mathrm{UC}}(u,v,x)=&\frac{2}{9}\left[\frac{v}{u}T_B(ux)T_\phi(vx)+\frac{u}{v}T_B(vx)T_\phi(ux)-\frac{1}{uv}T_B(ux)T_B(vx)\right]\\
=&\frac{6(u^2+v^2-3)}{u^3v^3x^4}\left(ux\cos\left(\frac{ux}{\sqrt{3}}\right)-\sqrt{3}\sin\left(\frac{ux}{\sqrt{3}}\right)\right)\\
&\times\left(vx\cos\left(\frac{vx}{\sqrt{3}}\right)-\sqrt{3}\sin\left(\frac{vx}{\sqrt{3}}\right)\right).
\end{split}
\end{equation}
Substituting Eq. \eqref{fuceq1} into Eq. \eqref{I_int}, we obtain
\begin{equation}
\label{iuceq1}
\begin{split}
I_{\mathrm{UC}}(u,v,x)=&\frac{3}{4u^3v^3x^4}\left[
-24\left(-ux\cos\frac{ux}{\sqrt{3}}+\sqrt{3}\sin\frac{ux}{\sqrt{3}}\right)\left(-vx\cos\frac{vx}{\sqrt{3}}+\sqrt{3}\sin\frac{vx}{\sqrt{3}}\right)\right.\\
&-4\left(
uv(u^2+v^2-3)x^3\sin x+6ux\cos\frac{ux}{\sqrt{3}}
\left(-vx \cos\frac{vx}{\sqrt{3}}+\sqrt{3}\sin\frac{vx}{\sqrt{3}}\right)\right.\\
&\left.-3\sin\frac{ux}{\sqrt{3}}\left(
-2\sqrt{3} vx\cos\frac{vx}{\sqrt{3}}+(6+(u^2+v^2-3)x^2)\sin\frac{vx}{\sqrt{3}}
\right)
\right)\\
&+(u^2+v^2-3)^2x^3\left(
\sin x
\left(
\mathrm{Ci}\left[\left(1+\frac{u-v}{\sqrt{3}}\right)x\right]+
\mathrm{Ci}\left[\left(1+\frac{v-u}{\sqrt{3}}\right)x\right]\right.\right.\\
&\left.\left.-\mathrm{Ci}\left[\left(1+\frac{u+v}{\sqrt{3}}\right)x\right]
-\mathrm{Ci}\left[\left|1-\frac{u+v}{\sqrt{3}}\right|x\right]
+\ln\left[
\left|\frac{3-(u+v)^2}{3-(u-v)^2}\right|
\right]
\right)\right.\\
&
+\cos x\left(
-\mathrm{Si}\left[\left(1+\frac{u-v}{\sqrt{3}}\right)x\right]-
\mathrm{Si}\left[\left(1+\frac{v-u}{\sqrt{3}}\right)x\right]+
\mathrm{Si}\left[\left(1-\frac{u+v}{\sqrt{3}}\right)x\right]\right.\\
&+\left.\left.\left.\mathrm{Si}\left[\left(1+\frac{u+v}{\sqrt{3}}\right)x\right]\right)\right)\right].
\end{split}
\end{equation}

To compare the result \eqref{iuceq1} with that from the Newtonian gauge by the gauge transformation,
we need the coordinate transformation
from the Newtonian gauge to the uniform curvature gauge
\begin{align}
\label{alphaunif}
  \alpha=&\frac{\phi_{\mathrm{N}}}{\mathcal{H}}=\frac23\zeta(\bm k)\frac1k T_\alpha(x),\\
  \beta=&0,
\end{align}
where $T_\alpha= xT_\text{N}$.

\begin{figure}
  \centering
  \includegraphics[width=0.6\textwidth]{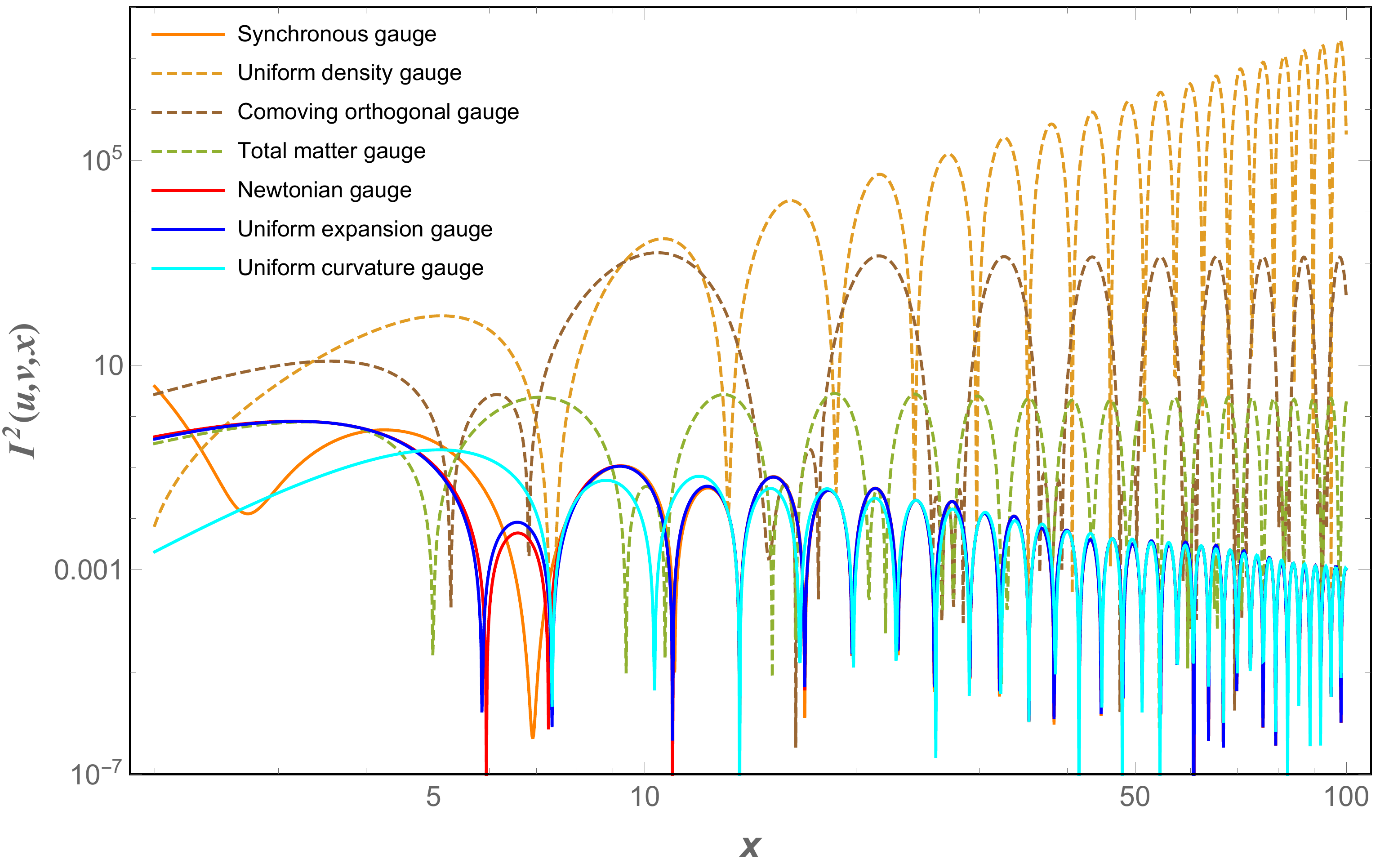}
  \caption{The evolution of the kernel $I^2(u,v,x)$ with $u=v=1$ in different gauges.}
\label{kenerluv1}
\end{figure}

Substituting these results into Eq. \eqref{ichi_newton}, we get
\begin{align}
\label{ichi_uni}
  I_\chi(u,v,x)=&-\frac{18}{u^3 v^3 x^4}\left[u v x^2 \cos \left(\frac{u x}{\sqrt{3}}\right) \cos \left(\frac{v x}{\sqrt{3}}\right)+3 \sin \left(\frac{u x}{\sqrt{3}}\right) \sin \left(\frac{v x}{\sqrt{3}}\right)\right.\notag\\
  &~\left.-\sqrt{3} v x \sin \left(\frac{u x}{\sqrt{3}}\right) \cos \left(\frac{v x}{\sqrt{3}}\right)-\sqrt{3} u x \cos \left(\frac{u x}{\sqrt{3}}\right) \sin \left(\frac{v x}{\sqrt{3}}\right)\right].
\end{align}
Therefore, we confirm that $I_{\mathrm{UC}}=I_{\mathrm N}+I_\chi$.
The evolution of $I^2_\text{UC}(u,v,x)$ with $u=v=1$ is shown in Fig. \ref{kenerluv1}.
Since $I_{\chi}(u,v,x)$ decays as $x^{-2}$ as $x\rightarrow\infty$, while $I_\mathrm{N}(u,v,x)$ decays as $x^{-1}$, the late time result in the
uniform curvature gauge is the same as that obtained in the Newtonian gauge, i.e.,  $I_{\mathrm{UC}}(u,v,x\rightarrow\infty)=I_{\mathrm N}(u,v,x\rightarrow\infty)$. This agrees with the conclusion in Refs. \cite{Yuan:2019fwv,Tomikawa:2019tvi}.
Notice that during the late time, the fractional energy density $\Omega_{\mathrm{GW}}\propto x^2I_{\mathrm{UC}}^2$. Thus, $\Omega_{\mathrm{GW}}$ becomes constant at late time in the uniform curvature gauge.

\subsection{Synchronous gauge}
Next, we consider the synchronous gauge, in which $\phi=B=0$, the equation for the transfer function $T_E$ is
\begin{equation}
\label{sync:teeq}
x^3T_{E}^{****}+5x^2T_E^{***}+\left(2+\frac{x^2}{3}\right)x T_E^{**}-\left(2-\frac{x^2}{3}\right)T_E^*=0,
\end{equation}
where the superscript ``*" on transfer functions denotes the derivative with respect to their arguments.
The general solution is
\begin{equation}
\label{sync:tesol}
T_E(x)=\mathcal{C}_1+\mathcal{C}_2\left(\mathrm{Ci}(x/\sqrt{3})-\frac{\sin(x/\sqrt{3})}{x/\sqrt{3}}\right)+\mathcal{C}_3\ln(x/\sqrt{3})+\mathcal{C}_4\left(\mathrm{Si}(x/\sqrt{3})+\frac{\cos(x/\sqrt{3})}{x/\sqrt{3}}\right),
\end{equation}
where $\mathcal{C}_i$ are integration constants.
Note that there are two gauge modes in Eq. \eqref{sync:tesol}
because of the residual gauge freedom in the synchronous gauge \cite{Press:1980is,Bucher:1999re,Bednarz:1984dn,Ma:1995ey}.
To identify these two gauge modes, during the radiation domination we take
the residual gauge transformation \cite{Press:1980is,Bucher:1999re}
\begin{equation}
\label{sync:transf}
\begin{split}
\alpha=&-\frac{\mathcal{C}_5}{x},\\
\beta=&-\mathcal{C}_5\ln x+\mathcal{C}_6.
\end{split}
\end{equation}
From the transformation \eqref{gauge:etransf}, we see that the constant
$\mathcal{C}_6$ term in $\beta$ contributes
to the integration constant $\mathcal{C}_1$ in Eq. \eqref{sync:tesol} and the $\mathcal{C}_5$ term in
$\beta$ contributes to the $\ln(x)$ term in Eq. \eqref{sync:tesol}.
Therefore, $\mathcal{C}_1$ and $\mathcal{C}_3$ terms in Eq. \eqref{sync:tesol} are just pure gauge modes.
Now we determine the remaining integration constants from the initial condition.
At the initial time $x=0$, $T_E(0)=0$, so we get $\mathcal{C}_4=0$,  $\mathcal{C}_1=-(1-\gamma_E)\mathcal{C}_2$ and $\mathcal{C}_3=-\mathcal{C}_2$, where the Euler gamma constant $\gamma_E\approx 0.577216$. Since $x\to 0$,
$\text{Ci}(x/\sqrt{3})-\ln(x/\sqrt{3})-\gamma_E\to 0$, so we need to add $\mathcal{C}_1$ and $\mathcal{C}_3$ terms to eliminate these gauge modes in $\text{Ci}(x)$ when $x\to 0$.
Finally we use the initial condition of the gauge invariant Bardeen potential to fix the constant $\mathcal{C}_2$.
The gauge invariant Bardeen potential in synchronous gauge is
\begin{align}
\Phi=-\mathcal{H}E^\prime-E^{\prime\prime},
\end{align}
so the transfer function $T_\Phi$ is
\begin{align}
T_\Phi=\frac{\mathcal{C}_2}{x^2}\left(\frac{\sin(x/\sqrt{3})}{x/\sqrt{3}}
-\cos(x/\sqrt{3})\right).
\end{align}
From the initial condition $T_\Phi(0)=1$,
we get $\mathcal{C}_2=9$, so $T_\Phi=T_\text{N}$ as expected because it is a gauge-invariant variable. Therefore, the transfer functions $T_E$ and $T_\psi$ are
\begin{equation}
\label{synte_psi}
\begin{split}
T_E(x)=&9\left[\mathcal{C}+\mathrm{Ci}(x/\sqrt{3})-\ln(x/\sqrt{3})-\frac{\sin(x/\sqrt{3})}{x/\sqrt{3}}
\right],\\
T_\psi(x)=&\frac{9}{x^2}\left(1-\cos(x/\sqrt{3})\right),
\end{split}
\end{equation}
where $\mathcal{C}=1-\gamma_E$. Recall that at late time $x\gg 1$,
$\mathcal{C}$ and $\ln(x/\sqrt{3})$ terms are gauge modes
and they are physical only at $x\ll 1$.
Therefore, at late time $x\gg1$, the transfer function is
\begin{equation}
\label{syn_te_late}
T_E(x)\approx 9\left[\text{Ci}(x/\sqrt{3})-\frac{\sin(x/\sqrt{3})}{x/\sqrt{3}}\right].
\end{equation}
Combining Eqs. \eqref{source} and \eqref{sfrel1}, we get
\begin{equation}\label{fsyn}
\begin{split}
f_{\mathrm{syn}}=&T_{\psi}(ux)T_{\psi}(vx)-\frac{2-u^2-v^2}{2uv}T_E^*(ux)T_E^*(vx)-\left(\frac{1-u^2-v^2}{2uv}\right)^2T_E(ux)T_E(vx)
\\&+\frac12\left[\frac{v}{u}T_E^*(ux)T_\psi^*(vx)+\frac{u}{v}T_E^*(vx)T_\psi^*(ux)\right]\\
&+x^2uvT_\psi^*(ux)T_\psi^*(vx)
+T_\psi(ux)T_E(vx)+T_\psi(vx)T_E(ux)\\
&+\frac{1}{u^2}T_E(ux)\left[v^2T_\psi^{**}(vx)+\frac{2v}{x}T_\psi^*(vx)+v^2T_\psi(vx)\right]+T_\psi(ux)T_E^{**}(vx)\\
&+\frac{1}{v^2}T_E(vx)\left[u^2T_\psi^{**}(ux)+\frac{2u}{x}T_\psi^*(ux)+u^2T_\psi(ux)\right]+T_\psi(vx)T_E^{**}(ux)\\
&+(1-u^2-v^2)\left[\frac{1}{v^2}T_\psi(ux)T_E(vx)
+\frac{1}{u^2}T_\psi(vx)T_E(ux)\right]
\\&
+2\left[\frac{1}{vx}T_\psi(ux)T_E^*(vx)+\frac{1}{ux}T_\psi(vx)T_E^*(ux)\right]\\
&-\frac{1-u^2-v^2}{4}\left(\frac{1}{u^2}T_E(ux)\left[T_E^{**}(vx)+\frac{2}{vx}T_E^*(vx)+T_E(vx)\right]\right.\\
&\left.+\frac{1}{v^2}T_E(vx)\left[T_E^{**}(ux)+\frac{2}{ux}T_E^*(ux)+T_E(ux)\right]\right).
\end{split}
\end{equation}
Since $I(u,v,x)$ depends on $f(u,v,x)$ linearly in Eq. \eqref{I_int}, we split $f_{\mathrm{Syn}}(u,v,x)$
as $f_{\mathrm{Syn}}=f_{\mathrm{N}}+\Delta f$ with $f_{\mathrm{N}}$ given by \eqref{fNewton}.
Substituting the result \eqref{fsyn} into Eq. \eqref{I_int}, we get $I_{\text{Syn}}=I_{\text{N}}+\Delta I_{\text{Syn}}(u,v,x)$ and
\begin{equation}
\label{syn_check}
\begin{split}
\Delta I_{\mathrm{Syn}}(u,v,x)=&\int_0^x\mathrm{d}\tilde x\frac{\tilde x}{x}kG_k(x;\tilde x) [f_\text{Syn}(u,v,\tilde x)-f_\text{N}(u,v,\tilde x)]\\
=&-\frac{9}{u^2v^2x^2}\left(
  (1-u^2-v^2)x^2\left[
\mathrm{Ci}\left(\frac{ux}{\sqrt{3}}\right)+\mathcal{C}-\ln\frac{ux}{\sqrt{3}}-\frac{\sin(ux/\sqrt{3})}{ux/\sqrt{3}}
  \right]\right.\\
  &\qquad \qquad \qquad \times\left[\mathrm{Ci}\left(\frac{vx}{\sqrt{3}}\right)+\mathcal{C}
  -\ln\frac{vx}{\sqrt{3}}-\frac{\sin(vx/\sqrt{3})}{vx/\sqrt{3}}\right]\\
  &\qquad \qquad +2\left[\frac{\sin(ux/\sqrt{3})}{ux/\sqrt{3}}-1\right]
  \left[\frac{\sin(vx/\sqrt{3})}{vx/\sqrt{3}}-1\right]\\
  &\qquad \qquad +4\left[-\mathrm{Ci}\left(\frac{ux}{\sqrt{3}}\right)
  -\mathcal{C}+\ln\frac{ux}{\sqrt{3}}+\frac{\sin(ux/\sqrt{3})}{ux/\sqrt{3}}\right]
  \left[1-\cos\frac{vx}{\sqrt{3}}\right]
  \\
  &\qquad \qquad \left.+4\left[-\mathrm{Ci}\left(\frac{vx}{\sqrt{3}}\right)-\mathcal{C}+\ln\frac{vx}{\sqrt{3}}
  +\frac{\sin(vx/\sqrt{3})}{vx/\sqrt{3}}\right]
  \left[1-\cos\frac{ux}{\sqrt{3}}\right]
  \right).
\end{split}
\end{equation}
We expect the result \eqref{syn_check} to equal $I_\chi(u,v,x)$ by the coordinate transformation from the Newtonian gauge to the synchronous gauge. As emphasized above,
in the synchronous gauge, we should include the contribution from $E$ in Eq. \eqref{source}. If the contribution from $E$ is not included then $I_\text{Syn}$ cannot be obtained from $I_\text{N}$ by the coordinate transformation from the Newtonian gauge to the synchronous gauge \cite{Yuan:2019fwv,Inomata:2019yww,DeLuca:2019ufz}.
To confirm our result \eqref{syn_check}, we now discuss the gauge transforms.
The coordinate transformation from the Newtonian gauge to the synchronous gauge is
\begin{equation}
\label{synalpha}
\begin{split}
  \alpha(\bm k,x)=&\frac23\zeta(\bm k)\frac1kT_\alpha(x),\\
  \beta(\bm k,x)=&\frac23\zeta(\bm k)\frac{1}{k^2}T_\beta(x),
\end{split}
\end{equation}
where the transfer functions are
\begin{equation}
\label{T_beta_syn}
\begin{split}
  T_\alpha(x)=&\frac{9}{x}\left[\frac{\sin(x/\sqrt{3})}{x/\sqrt{3}}-1\right],\\
  T_\beta(x)=&9\left[\mathcal{C}+\text{Ci}\left(\frac{x}{\sqrt{3}}\right)-\ln\left(\frac{x}{\sqrt{3}}\right)-
  \frac{\sin(x/\sqrt{3})}{x/\sqrt{3}}\right].
\end{split}
\end{equation}
Substituting Eq. \eqref{T_beta_syn} into Eq. \eqref{ichi_newton},  we find
$I_{\chi}^{\mathrm{Syn}}(u,v,x)=\Delta I_{\mathrm{Syn}}(u,v,x)$
and confirm that $I_{\text{Syn}}(u,v,x)=I_{\text{N}}(u,v,x)+I_{\chi}(u,v,x)$.
The evolution of $I^2_\text{Syn}(u,v,x)$ with $u=v=1$ is shown in Fig. \ref{kenerluv1}.
As discussed above, at late time with $x\gg 1$ the growing mode $\ln(ux/\sqrt{3})\ln(vx/\sqrt{3})$ in $I_\chi(u,v,x)$ is a gauge mode.
Dropping the gauge terms, we find that the contribution of $E$ is negligible.
Therefore, at late time $I_{\text{Syn}}=I_\text{N}$ and
the result on the energy density of SIGWs
is the same in both the Newtonian gauge and the synchronous gauge,
as found in \cite{Yuan:2019fwv,Inomata:2019yww,DeLuca:2019ufz}.
Although at late time the contribution of $E$ is negligible
and it does not affect the obtained energy density of SIGWs,
we still need to include $E$ in the calculation so that the covariance
of $h_{ij}$ is guaranteed and the relation between $I_\text{Syn}$
and $I_\text{N}$ under the coordinate transformation is retained. In the synchronous gauge the subtlety arises in determining when to eliminate the gauge modes in $\text{Ci}(x/\sqrt{3})-\ln(x/\sqrt{3})+\mathcal{C}$. Accordingly, this gauge is not a particularly good choice for calculating the production of SIGWs.

\subsection{Comoving gauge (total matter gauge)}

In the comoving gauge (also referred as the total matter gauge \cite{Malik:2008im}),
$\delta V=E=0$, and the transfer functions are
\begin{equation}
\label{transfer_psi_TM}
\begin{split}
T_\psi(x)=&\frac32\frac{\sin(x/\sqrt{3})}{x/\sqrt{3}},\\
T_B(x)=&\frac{3}{2x^2}\left[
6x\cos(x/\sqrt{3})+\sqrt{3}(x^2-6)\sin(x/\sqrt{3})
\right],\\
T_\phi(x)=&\frac32\left[
\frac{\sin(x/\sqrt{3})}{x/\sqrt{3}}-\cos(x/\sqrt{3})
\right].
\end{split}
\end{equation}
As expected, on the superhorizon scales, $\psi(\bm k,0)=\zeta(\bm k)$,
so $T_\psi(0)=3/2$ and at late time with $x\gg 1$, the perturbation $\psi$ decays as \cite{Tomikawa:2019tvi}
\begin{equation}
\label{tmpsi}
\psi(\bm k,\eta)=\psi(\bm k,0)\frac{\sin(x/\sqrt{3})}{x/\sqrt{3}}.
\end{equation}
However, the perturbations $B$ and $\phi$ do not decay at late time with $x\gg 1$ \cite{Tomikawa:2019tvi},
they will induce SIGWs continuously.
Combining Eqs. \eqref{source}, \eqref{sfrel1} and \eqref{transfer_psi_TM}, we have \cite{Tomikawa:2019tvi}
\begin{equation}
\label{tmfeq}
\begin{split}
f_{\mathrm{TM}}(u,v,x)
=&
\frac{1}{2u^3v^3x^4}\left[\vphantom{\frac12}\right.
2ux\cos\frac{ux}{\sqrt{3}}\left(\vphantom{\frac12}\right.
-vx(18-12v^2+u^2(v^2x^2-12))\cos\frac{vx}{\sqrt{3}}\\&+\sqrt{3}(18+v^4x^2-3v^2(4+x^2)
+2u^2(v^2x^2-6))\sin\frac{vx}{\sqrt{3}}
\left.\vphantom{\frac12}\right)\\&
+\sin\frac{ux}{\sqrt{3}}\left(\vphantom{\frac12}\right.
2\sqrt{3} vx(18-12v^2+u^4x^2+u^2(-12+(2v^2-3)x^2))\cos\frac{vx}{\sqrt{3}}\\
&+(u^4x^2(v^2x^2-6)+u^2(72-6(2v^2-3)x^2+v^2(v^2-3)x^4)\\&-6(18+v^4x^2-3v^2(4+x^2)))\sin\frac{vx}{\sqrt{3}}
\left.\vphantom{\frac12}\right)\left.\vphantom{\frac12}
\right].
\end{split}
\end{equation}
Substituting Eq. \eqref{tmfeq} into Eq. \eqref{I_int}, we get
\begin{equation}
\label{tmieq}
\begin{split}
I_{\mathrm{TM}}(u,v,x)=&
\frac{1}{4u^3v^3x^4}\left(
-2\left(
6ux\cos\frac{ux}{\sqrt{3}}+\sqrt{3}(u^2x^2-6)\sin\frac{ux}{\sqrt{3}}
\right)\right.\\
&\times\left(
6vx\cos\frac{vx}{\sqrt{3}}+\sqrt{3}(v^2x^2-6)\sin\frac{vx}{\sqrt{3}}
\right)\\
&-12\left[
uv(u^2+v^2-3)x^3\sin x+6ux\cos\frac{ux}{\sqrt{3}}\left(
-vx\cos\frac{vx}{\sqrt{3}}+\sqrt{3}\sin\frac{vx}{\sqrt{3}}
\right)\right.\\
&\left.-3\sin\frac{ux}{\sqrt{3}}\left(
-2\sqrt{3}vx\cos\frac{vx}{\sqrt{3}}+[6+(u^2+v^2-3)^2 x^2]\sin\frac{vx}{\sqrt{3}}\right)\right]\\
&+3(u^2+v^2-3)x^3\left[
\sin x\left(\mathrm{Ci}\left[\left(1+\frac{u-v}{\sqrt{3}}\right)x\right]+
\mathrm{Ci}\left[\left(1+\frac{v-u}{\sqrt{3}}\right)x\right]\right.\right.\\
&\left.-\mathrm{Ci}\left[\left(1+\frac{u+v}{\sqrt{3}}\right)x\right]
-\mathrm{Ci}\left[\left|1-\frac{u+v}{\sqrt{3}}\right|x\right]
+\ln\left[\left|\frac{3-(u+v)^2}{3-(u-v)^2}\right|\right]\right)\\
&+\cos x\left(-\mathrm{Si}\left[\left(1+\frac{u-v}{\sqrt{3}}\right)x\right]-
\mathrm{Si}\left[\left(1+\frac{v-u}{\sqrt{3}}\right)x\right]\right.\\
&\left.\left.\left.+\mathrm{Si}\left[\left(1-\frac{u+v}{\sqrt{3}}\right)x\right]+
\mathrm{Si}\left[\left(1+\frac{u+v}{\sqrt{3}}\right)x\right]\right)\right]\right).
\end{split}
\end{equation}
At the late time, $I_{\mathrm{TM}}(u,v,x\to\infty)$ approaches to a constant.

From the Newtonian gauge to the comoving gauge, the
transfer functions for the coordinate transformation are
\begin{align}\label{alphacomov}
  \alpha=&\frac{\mathcal{H}\phi_{\mathrm N}+\phi_{\mathrm N}^\prime}{\mathcal{H}'-\mathcal{H}^2}=\frac23\zeta(\bm k)\frac1kT_\alpha(x),\\
  \beta=&0,
\end{align}
where
\begin{align}\label{transferTM}
  T_\alpha(x)=&-\frac12(xT_\mathrm{N}(x)+x^2T_{\mathrm N}^*(x))\notag\\
  =&-\frac{3}{2x^2}\left[
6x\cos(x/\sqrt{3})+\sqrt{3}(x^2-6)\sin(x/\sqrt{3})
\right].
\end{align}
Substituting Eq. \eqref{transferTM} into  Eq. \eqref{ichi_newton}, we get
\begin{align}
\label{Ichi_tm}
  I_\chi(u,v,x)=-\frac{3}{2u^3 v^3 x^4}&\left[ \left(u^2 x^2-6\right) \sin \left(\frac{u x}{\sqrt{3}}\right) \left(\left(v^2 x^2-6\right) \sin \left(\frac{v x}{\sqrt{3}}\right)+2 \sqrt{3} v x \cos \left(\frac{v x}{\sqrt{3}}\right)\right)\right.\notag\\
  &~\left.
  +2 u x \cos \left(\frac{u x}{\sqrt{3}}\right) \left(\sqrt{3} \left(v^2 x^2-6\right) \sin \left(\frac{v x}{\sqrt{3}}\right)+6 v x \cos \left(\frac{v x}{\sqrt{3}}\right)\right)
  \right].
\end{align}
Again we confirm that $I_{\mathrm{TM}}(u,v,x)=I_\mathrm{N}(u,v,x)+I_\chi(u,v,x)$.
The evolution of $I^2_\text{TM}(u,v,x)$ with $u=v=1$ is shown in Fig. \ref{kenerluv1}.
It is obvious that at late time, the constant term in $I_\chi(u,v,x)$ dominates over $I_\text{N}$,
so $I_{\mathrm{TM}}(u,v,x\rightarrow\infty)$ approaches a constant.
In this gauge, the perturbations $B$ and $\phi$ do not decay as $x\to \infty$ and
they induce SIGWs continuously, so
$\Omega_{\mathrm{GW}}$ for SIGWs grows as $x^2$.
This result agrees with Ref. \cite{Tomikawa:2019tvi}.

\subsection{Comoving orthogonal gauge}

Let us now focus on the comoving orthogonal gauge, $\delta V=B=0$ \cite{Malik:2008im}.
In this gauge, there remains a residual coordinate
transformation with $\beta=\mathcal{C}$
which corresponds to the arbitrary choice of the origin of the spatial coordinates.
The variable $\alpha$ for the time coordinate transformation from the Newtonian gauge
to this gauge is the same as that
from the Newtonian gauge to the total matter gauge.
The variable $\beta$ for the spatial coordinate transformation
from the Newtonian gauge to this gauge is
\begin{equation}
\label{cogbeta}
  \beta\left(\bm{k},\eta\right)=\frac{2}{3}\zeta\left(\bm{k}\right)
\frac{1}{k^2}T_{\beta}\left(x\right),
\end{equation}
where
\begin{equation}
\label{cotransfer}
  T_\beta(x) =-9\left[\cos\left(x/\sqrt{3}\right)
-\frac{2\sqrt{3}\sin\left(x/\sqrt{3}\right)}{x}+\mathcal{C}\right].
\end{equation}
We may choose $\mathcal{C}=1$ so that ${T_{\beta}(x=0)=0}$.
At late time, $x\gg 1$, the last constant $\mathcal{C}$ term is a pure gauge mode, and both $T_\alpha$ and $T_\beta$ do not decay.
Substituting Eq. \eqref{cotransfer} into  Eq. \eqref{ichi_newton}, we get
\begin{equation}
\begin{split}
I_{\chi}\left(u,v,x\right)=&\frac{3}{4u^3v^3x^4}\left[
3\mathcal{C}^2uv(u^2+v^2-1)x^4-2\sqrt{3}\mathcal{C}v(5u^2+3v^2-3)x^3\sin\frac{ux}{\sqrt{3}}\right.\\
&
-2\sqrt{3}\mathcal{C}u(3u^2+5v^2-3)x^3\sin\frac{vx}{\sqrt{3}}\\
&
-2[36-18(2u^2+2v^2-1)x^2+u^2v^2x^4]\sin\frac{ux}{\sqrt{3}}\sin\frac{vx}{\sqrt{3}}
\\
&+3uvx^2[-8+(u^2+v^2-1)x^2]\cos\frac{ux}{\sqrt{3}}\cos\frac{vx}{\sqrt{3}}
\\
&+vx\cos\frac{vx}{\sqrt{3}}\left(3\mathcal{C}u(u^2+v^2-1)x^3
-2\sqrt{3}[-12+(7u^2+3v^2-3)x^2]\sin\frac{ux}{\sqrt{3}}\right)\\
&\left.+ux\cos\frac{ux}{\sqrt{3}}
\left(
3\mathcal{C}v(u^2+v^2-1)x^3
-2\sqrt{3}[-12+(3u^2+7v^2-3)x^2]\sin\frac{vx}{\sqrt{3}}\right)
\right],
\end{split}
\end{equation}
and the analytic expression for the kernel $I_{\mathrm{CO}}$ in the comoving orthogonal gauge is $I_{\mathrm{CO}}=I_\mathrm{N}+I_\chi$.
The evolution of $I^2_\text{CO}(u,v,x)$ with $u=v=1$ is shown in Fig. \ref{kenerluv1}.
At late time, even after dropping the gauge mode,
$I_\chi$ still approaches to a constant because $\phi$ and $E$ do not decay,
so $\Omega_{\mathrm{GW}}$ for SIGWs grows as $x^2$.

\subsection{Uniform density gauge}

\begin{figure}
  \centering
  \includegraphics[width=0.6\textwidth]{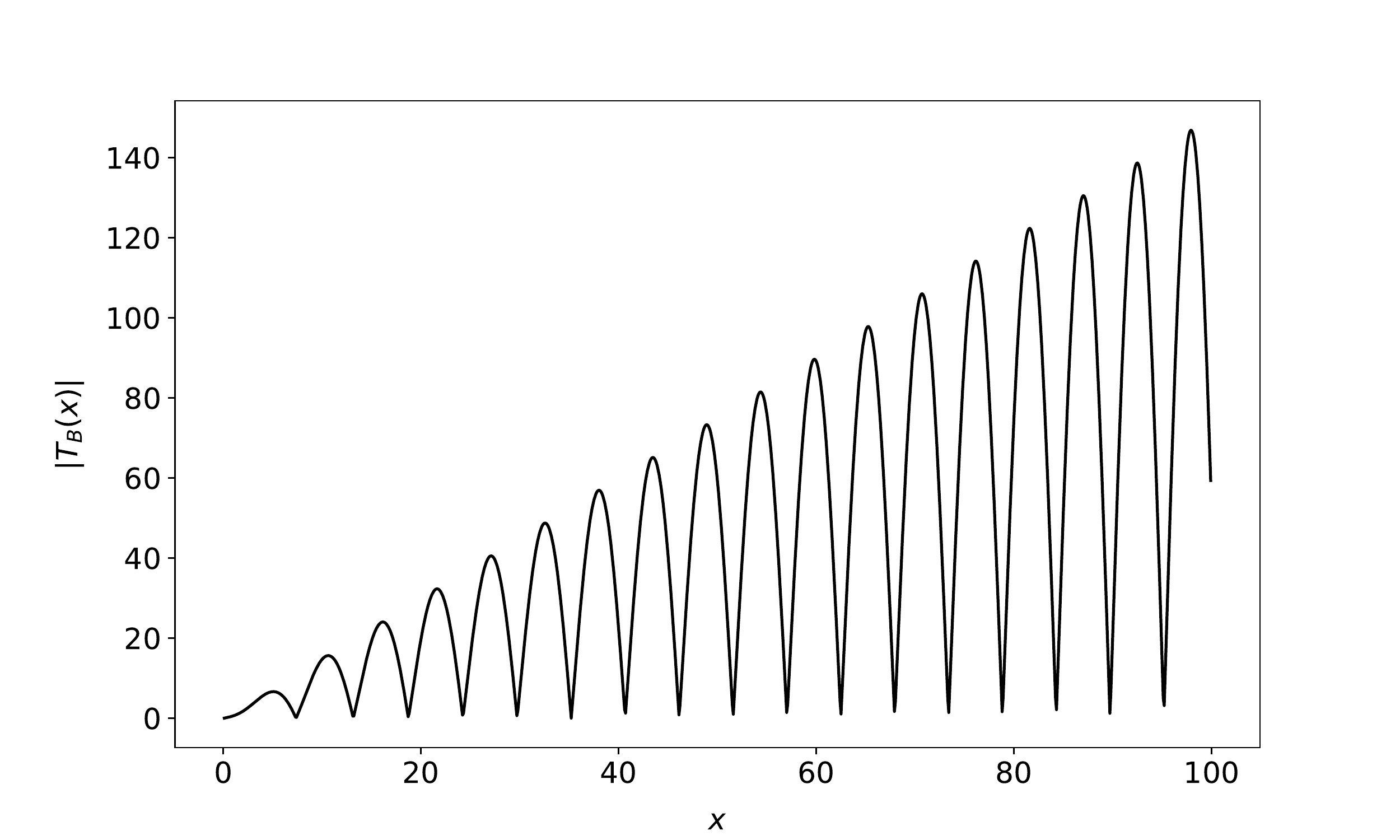}
  \caption{The behavior of the transfer function $T_B$ in the uniform density gauge.}
\label{fig2}
\end{figure}

The uniform density gauge is defined as $\delta\rho=E=0$.
The coordinate transformation
from the Newtonian gauge to this gauge is
\begin{equation}
\begin{split}
\label{UDalpha}
  \alpha&=-\frac{\delta\rho_{\text{N}}}{\rho'_0},\\
  \beta&=0.
\end{split}
\end{equation}
Using the first-order perturbation equation and
the background equation, we obtain
\begin{equation}
\label{udalpha2}
\begin{split}
  \alpha=&\frac{\mathcal{H}\phi_{\mathrm N}+\phi_{\mathrm N}^\prime}{\mathcal{H}'-\mathcal{H}^2}+\frac{k^2\phi_\mathrm{N}}{3\mathcal{H}(\mathcal{H}^\prime-\mathcal{H}^2)}\\
  =&\frac23\zeta(\bm k)\frac1kT_\alpha(x),
\end{split}
\end{equation}
where
\begin{align}
\label{transferfuncUD}
  T_\alpha(x)=\frac{3}{2x^2}\left[x(x^2-6)\cos(x/\sqrt{3})-2\sqrt{3}(x^2-3)\sin(x/\sqrt{3})
  \right].
\end{align}
At late time, the first term in Eq. \eqref{transferfuncUD} grows and we
may wonder whether it violates the condition of infinitesimal coordinate transformation. To check this, we consider the dimensionless variable $\alpha/\eta$,
\begin{equation}
\label{udginft}
\frac{\alpha}{\eta}\to \zeta(k)\cos(x/\sqrt{3}),\quad x\to \infty.
\end{equation}
Therefore, the condition of the infinitesimal coordinate transformation is still satisfied at late time.
Substituting Eq. \eqref{transferfuncUD} into  Eq. \eqref{ichi_newton},
we get
\begin{equation}
\label{IalphaUD}
\begin{split}
  I_{\chi}(u,v,x)=&\frac{1}{4 u^3 v^3 x^4}\left[
 2 u x \left(u^2 x^2-6\right) \cos \left(\frac{u x}{\sqrt{3}}\right)\left(v x \left(v^2 x^2-6\right) \cos \left(\frac{v x}{\sqrt{3}}\right)\right.\right.\\
 &\left.-2 \sqrt{3} \left(v^2 x^2-3\right) \sin \left(\frac{v x}{\sqrt{3}}\right)\right)
 -4 \left(u^2 x^2-3\right) \sin \left(\frac{u x}{\sqrt{3}}\right)\\
 &\times\left.\left(\sqrt{3} v x \left(v^2 x^2-6\right) \cos \left(\frac{v x}{\sqrt{3}}\right)-6 \left(v^2 x^2-3\right) \sin \left(\frac{v x}{\sqrt{3}}\right)\right)
 \right],
 \end{split}
\end{equation}
and the analytic expression for the kernel in
the uniform density gauge $I_{\mathrm{UD}}=I_{\mathrm N}+I_{\chi}$.
The evolution of $I^2_\text{UD}(u,v,x)$ with $u=v=1$ is shown in Fig. \ref{kenerluv1}.
At late time, $I_{\chi}$ grows as $x^{2}$, so $\Omega_{\mathrm{GW}}^{\mathrm{UD}}\sim x^6$.
In this gauge, $\phi$ and $\psi$ do not decay, and the variable $B=-\alpha$ even grows with $x$ as $x\to \infty$. If we assume $\zeta(k)\sim 0.01$, then $kB$ approaches order 1 when $T_B$ is about 100 at $x\sim 100$ as shown in Fig. \ref{fig2}, so the linear perturbation breaks down and the above calculation cannot be  applied. Thus, we need to be careful about the calculation of $\Omega_{\text{GW}}$ in this gauge.

\subsection{Uniform expansion gauge}
Let us finally consider the uniform expansion gauge,
$3(\mathcal{H}\phi+\psi^\prime)+k^2\sigma=0$ and $E=0$ \cite{Hwang:2017oxa}. From the Newtonian gauge to this gauge, the coordinate transformation is
\begin{equation}
\label{alphauniexp}
\begin{split}
  \alpha=&\frac23\zeta(\bm k)\frac1kT_\alpha(x),\\
  \beta=&0,
\end{split}
\end{equation}
where
\begin{align}
\label{transferfuncuniexp}
  T_\alpha(x)=&\frac{-3\left(x T_{\mathrm N}(x)+x^2T_{\mathrm N}^*(x)\right)}{6+x^2}
  \notag\\=&-\frac{9}{x^2(6+x^2)}\left[6x\cos(x/\sqrt{3})+\sqrt{3}(x^2-6)\sin(x/\sqrt{3})
  \right].
\end{align}
Substituting Eq. \eqref{transferfuncuniexp} into  Eq. \eqref{ichi_newton}, we get
\begin{equation}
\label{IalphaUniexp}
\begin{split}
  I_{\chi}(u,v,x)=&\frac{54}{u^3 v^3 x^4 \left(u^2 x^2+6\right) \left(v^2 x^2+6\right)}\left[
  u^2 v^2 x^4 \sin \left(\frac{u x}{\sqrt{3}}\right) \sin \left(\frac{v x}{\sqrt{3}}\right)\right.\\
  &
  +2 \sqrt{3} u^2 v x^3 \sin \left(\frac{u x}{\sqrt{3}}\right) \cos \left(\frac{v x}{\sqrt{3}}\right)-6 u^2 x^2 \sin \left(\frac{u x}{\sqrt{3}}\right) \sin \left(\frac{v x}{\sqrt{3}}\right)\\
  &
  +2 \sqrt{3} u v^2 x^3 \cos \left(\frac{u x}{\sqrt{3}}\right) \sin \left(\frac{v x}{\sqrt{3}}\right)-6 v^2 x^2 \sin \left(\frac{u x}{\sqrt{3}}\right) \sin \left(\frac{v x}{\sqrt{3}}\right)\\
  &+12 u v x^2 \cos \left(\frac{u x}{\sqrt{3}}\right) \cos \left(\frac{v x}{\sqrt{3}}\right)+36 \sin \left(\frac{u x}{\sqrt{3}}\right) \sin \left(\frac{v x}{\sqrt{3}}\right)\\
  &\left.-12 \sqrt{3} v x \sin \left(\frac{u x}{\sqrt{3}}\right) \cos \left(\frac{v x}{\sqrt{3}}\right)
  -12 \sqrt{3} u x \cos \left(\frac{u x}{\sqrt{3}}\right) \sin \left(\frac{v x}{\sqrt{3}}\right)\right].
\end{split}
\end{equation}
So the analytic expression for the kernel $I_{\mathrm{UE}}$ in the uniform
expansion gauge is $I_{\mathrm{UE}}(u,v,x)=I_\mathrm{N}+I_{\chi}$,
and it decays as $x^{-4}$ when $x\rightarrow\infty$,
indicating that $\Omega_{\mathrm{GW}}^{\mathrm{UE}}$ approaches $\Omega_{\mathrm{GW}}^{\mathrm{N}}$ at late time.
The evolution of $I^2_\text{UE}(u,v,x)$ with $u=v=1$ is shown in Fig. \ref{kenerluv1}.

\section{Conclusion}
\label{sec4}

We derive the general formula valid in any gauge for the calculation of SIGWs.
In particular, we provide the prescription to use the result in the Newtonian gauge to obtain SIGWs in several other gauges by the coordinate transformation from the Newtonian gauge to the other gauges, and also provide the general expression for the
kernel function $I_\chi(u,v,x)$. Besides, we directly derive the kernel functions in the uniform curvature gauge, the synchronous gauge and
the total matter gauge, and confirm that they are the same
as those obtained by the coordinate transformation
from the Newtonian gauge to the other gauges.
With the general kernel function $I_\chi(u,v,x)$
and the result of $\Omega_{\text{GW}}$ in the Newtonian gauge,
we derive the results of $\Omega_{\text{GW}}$ in
the comoving orthogonal gauge, the uniform density gauge
and the uniform expansion gauge by the coordinate
transformation form the Newtonian gauge to these gauges.
The Newtonian gauge, the uniform curvature gauge,
the synchronous gauge and the uniform expansion gauge have
the same result on $\Omega_{\text{GW}}$.

We also identify the two gauge modes in the synchronous gauge, which lead to the growing
of the kernel function, and we find that $\Omega_{\text{GW}}$
in the synchronous gauge is the same as that in the Newtonian gauge
after eliminating the gauge modes.
Although the contribution of the perturbation $E$
is negligible after eliminating the gauge modes at late time,
and $E$ does not affect the final result on the energy
density of SIGWs, its contribution cannot be neglected.
Otherwise the relationship
between the results for $\Omega_{\text{GW}}$
in the synchronous gauge and  the other gauges
under the gauge transformation is not satisfied.
Since the constant term $\mathcal{C}$ and the $\ln(x)$ term
are gauge modes only at late time with $x\gg 1$,
they are physical modes at early time with $x\ll 1$;
thus, determining when to drop these terms is quite subtle. Thus, in our view the synchronous gauge is not a good choice for the
calculation of the production of SIGWs.

Finally, in the total matter gauge and the comoving orthogonal gauge,
the perturbation $\phi$ does not decay at late time,
so as $x\gg 1$ the energy density of SIGWs
in both gauges increases as $x^2$.
In the uniform density gauge, the perturbation $B$
grows as $x$ at late time, so as $x\gg 1$
the energy density of SIGWs
in this gauge increase as $x^6$.
Of course, we need to be careful about this result because
the perturbation theory breaks down due to the growth of $B$.
The reason for the gauge dependence of the energy density
of SIGWs and the issue of the observable need to be further studied.

\begin{acknowledgments}
Y.L. would like to thank Takahiro Terada for useful discussion.
This research was supported in part by the National Natural Science Foundation of
China under Grant No. 11875136 and the Major Program of the National Natural Science
Foundation of China under Grant No. 11690021.
\end{acknowledgments}

\appendix
\section{The perturbation and gauge transformation}
\label{app.A}

In this appendix we provide some details about the perturbation and gauge transformation for the SIGWs.

The energy-momentum tensor of a perfect fluid is
\begin{align}\label{stresstensor}
  T_{\mu\nu}=(\rho+P)U_\mu U_\nu+Pg_{\mu\nu}+\Pi_{\mu\nu},
\end{align}
where the background anisotropic stress ${\Pi}_{0\mu\nu}$ is assumed to be zero.
The first-order perturbations of the velocity $U_\mu$,
the energy density, the pressure
and the anisotropic stress are $\delta U_\mu$, $\delta\rho$, $\delta P$ and $\delta\Pi_{ij}$, respectively. The first-order velocity perturbation
$\delta U_\mu$ is decomposed via $\delta U_\mu=a[\delta V_0,\delta V_{,i}+\delta V_i]$ with $\delta V_{i,i}=0$.

For flat Friedmann-Robertson-Walker space-time, the background cosmological equations imply that
\begin{equation}
\label{backeqs}
\begin{split}
  \mathcal{H}^2=&\frac{8\pi G}{3}a^2{\rho_0},\\
  \mathcal{H}^\prime=&-\frac{4\pi G}{3}a^2(\rho_0+3P_0).
\end{split}
\end{equation}
For the discussion of the perturbed equation, we also write the
above Friedmann equations as
\begin{equation}
\label{backeqs_2}
\begin{split}
  {\rho}_0+P_0=\frac{\mathcal{H}^2-\mathcal{H}^\prime}{4\pi G a^2},\\
  P_0=-\frac{\mathcal{H}^2+2\mathcal{H}^\prime}{8\pi Ga^2}.
\end{split}
\end{equation}
In the absence of the anisotropic stress,
the first-order perturbed cosmological equations are

\begin{gather}
 \label{1st_eq_1}
 3\mathcal{H}(\psi^\prime+\mathcal{H}\phi)-\nabla^2(\psi+\mathcal{H}\sigma)
  =-4\pi Ga^2\delta\rho,\\
\label{1st_eq_2}
\psi^\prime+\mathcal{H}\phi=-4\pi Ga^2(\rho_0+P_0)\delta V,\\
\label{1st_eq_3}
\sigma'+2\mathcal{H}\sigma+\psi-\phi=0,\\
\label{1st_eq_4}
\psi^{\prime\prime}+2\mathcal{H}
\psi^\prime+\mathcal{H}\phi^\prime+(2\mathcal{H}^\prime+\mathcal{H}^2)\phi=4\pi Ga^2\delta P.
\end{gather}

Under the infinitesimal coordinate transformation
$x^\mu\rightarrow \tilde{x}^\mu=x^\mu+\epsilon^\mu(x)$ with
$\epsilon^\mu=[\alpha,\delta^{ij}\partial_j\beta]$,
the scalar parts of the perturbations transform as
\begin{gather}
\tilde{\phi}=\phi + \mathcal{H}\alpha + \alpha^\prime,\\
\tilde{\psi}=\psi -\mathcal{H}\alpha,\\
\tilde{B}=B -\alpha +\beta',\\
\label{gauge:etransf}
\tilde{E}=E+\beta,\\
\tilde{\sigma}=\sigma+\alpha,\\
\delta\tilde{\rho}=\delta\rho+{\rho}_0^\prime\alpha,\\
\delta\tilde{P}=\delta P+P_0^\prime\alpha,\\
\delta\tilde{V}=\delta V-\alpha,\\
\delta\tilde{\Pi}=\delta\Pi,
\end{gather}
where $\Pi$ is the scalar part of the anisotropic stress.
Using the above gauge transformation, we obtain two gauge-invariant Bardeen potentials \cite{Bardeen:1980kt}
\begin{gather}
\label{bardeenphi}
\Phi=\phi-\mathcal{H}\sigma-\sigma^\prime,\\
\label{bardeenpsi}
\Psi= \psi+\mathcal{H}\sigma.
\end{gather}
For the SIGWs, under the infinitesimal coordinate transformation,
we have $h_{ij}^{\mathrm{TT}}\to h_{ij}^{\mathrm{TT}}+\chi^{\mathrm{TT}}_{ij}$, and
\begin{equation}
\label{hgaugetrans}
\begin{split}
\chi_{ij}=&2\left[\left(\mathcal{H}^2+\frac{a^{''}}{a}\right)\alpha^{2}
+\mathcal{H}\left(\alpha\alpha'+\alpha_{,k}\epsilon^{k}\right)\right]\delta_{ij}\\
&+4\left[\alpha\left(C'_{ij}+2\mathcal{H}C_{ij}\right)+C_{ij,k}\epsilon^k+
C_{ik}\epsilon^k_{,j}+C_{jk}\epsilon^k_{,i}\right]\\
&+2\left(B_{i}\alpha_{,j}+B_{j}\alpha_{,i}\right)
+4\mathcal{H}\alpha\left(\epsilon_{i,j}+\epsilon_{j,i}\right)
-2\alpha_{,i}\alpha_{,j}+2\epsilon_{k,i}\epsilon^{k}_{,j}+
\alpha\left(\epsilon'_{i,j}+\epsilon'_{j,i}\right)\\
&+\left(\epsilon_{i,jk}+\epsilon_{j,ik}\right)\epsilon^k +\epsilon_{i,k}\epsilon^{k}_{,j}+\epsilon_{j,k}\epsilon^{k}_{,i}+\epsilon'_{i}\alpha_{,j}+\epsilon'_{j}\alpha_{,i},
\end{split}
\end{equation}
where $C_{ij}=-\psi\delta_{ij}+E_{,ij}$.


\begin{thebibliography}{62}%
\makeatletter
\providecommand \@ifxundefined [1]{%
 \@ifx{#1\undefined}
}%
\providecommand \@ifnum [1]{%
 \ifnum #1\expandafter \@firstoftwo
 \else \expandafter \@secondoftwo
 \fi
}%
\providecommand \@ifx [1]{%
 \ifx #1\expandafter \@firstoftwo
 \else \expandafter \@secondoftwo
 \fi
}%
\providecommand \natexlab [1]{#1}%
\providecommand \enquote  [1]{``#1''}%
\providecommand \bibnamefont  [1]{#1}%
\providecommand \bibfnamefont [1]{#1}%
\providecommand \citenamefont [1]{#1}%
\providecommand \href@noop [0]{\@secondoftwo}%
\providecommand \href [0]{\begingroup \@sanitize@url \@href}%
\providecommand \@href[1]{\@@startlink{#1}\@@href}%
\providecommand \@@href[1]{\endgroup#1\@@endlink}%
\providecommand \@sanitize@url [0]{\catcode `\\12\catcode `\$12\catcode
  `\&12\catcode `\#12\catcode `\^12\catcode `\_12\catcode `\%12\relax}%
\providecommand \@@startlink[1]{}%
\providecommand \@@endlink[0]{}%
\providecommand \url  [0]{\begingroup\@sanitize@url \@url }%
\providecommand \@url [1]{\endgroup\@href {#1}{\urlprefix }}%
\providecommand \urlprefix  [0]{URL }%
\providecommand \Eprint [0]{\href }%
\providecommand \doibase [0]{https://doi.org/}%
\providecommand \selectlanguage [0]{\@gobble}%
\providecommand \bibinfo  [0]{\@secondoftwo}%
\providecommand \bibfield  [0]{\@secondoftwo}%
\providecommand \translation [1]{[#1]}%
\providecommand \BibitemOpen [0]{}%
\providecommand \bibitemStop [0]{}%
\providecommand \bibitemNoStop [0]{.\EOS\space}%
\providecommand \EOS [0]{\spacefactor3000\relax}%
\providecommand \BibitemShut  [1]{\csname bibitem#1\endcsname}%
\let\auto@bib@innerbib\@empty
\bibitem [{\citenamefont {Abbott}\ \emph
  {et~al.}(2016{\natexlab{a}})\citenamefont {Abbott} \emph
  {et~al.}}]{Abbott:2016blz}%
  \BibitemOpen
  \bibfield  {author} {\bibinfo {author} {\bibfnamefont {B.~P.}\ \bibnamefont
  {Abbott}} \emph {et~al.} (\bibinfo {collaboration} {LIGO Scientific and Virgo
  Collaborations}),\ }\bibinfo {title} {{Observation of Gravitational Waves
  from a Binary Black Hole Merger}},\ \href
  {https://doi.org/10.1103/PhysRevLett.116.061102} {\bibfield  {journal}
  {\bibinfo  {journal} {Phys. Rev. Lett.}\ }\textbf {\bibinfo {volume} {116}},\
  \bibinfo {pages} {061102} (\bibinfo {year} {2016}{\natexlab{a}})},\ \Eprint
  {https://arxiv.org/abs/1602.03837} {arXiv:1602.03837} \BibitemShut {NoStop}%
\bibitem [{\citenamefont {Abbott}\ \emph
  {et~al.}(2016{\natexlab{b}})\citenamefont {Abbott} \emph
  {et~al.}}]{Abbott:2016nmj}%
  \BibitemOpen
  \bibfield  {author} {\bibinfo {author} {\bibfnamefont {B.~P.}\ \bibnamefont
  {Abbott}} \emph {et~al.} (\bibinfo {collaboration} {Virgo, LIGO
  Scientific}),\ }\bibinfo {title} {{GW151226: Observation of Gravitational
  Waves from a 22-Solar-Mass Binary Black Hole Coalescence}},\ \href
  {https://doi.org/10.1103/PhysRevLett.116.241103} {\bibfield  {journal}
  {\bibinfo  {journal} {Phys. Rev. Lett.}\ }\textbf {\bibinfo {volume} {116}},\
  \bibinfo {pages} {241103} (\bibinfo {year} {2016}{\natexlab{b}})},\ \Eprint
  {https://arxiv.org/abs/1606.04855} {arXiv:1606.04855} \BibitemShut {NoStop}%
\bibitem [{\citenamefont {Abbott}\ \emph
  {et~al.}(2017{\natexlab{a}})\citenamefont {Abbott} \emph
  {et~al.}}]{Abbott:2017vtc}%
  \BibitemOpen
  \bibfield  {author} {\bibinfo {author} {\bibfnamefont {B.~P.}\ \bibnamefont
  {Abbott}} \emph {et~al.} (\bibinfo {collaboration} {VIRGO, LIGO
  Scientific}),\ }\bibinfo {title} {{GW170104: Observation of a 50-Solar-Mass
  Binary Black Hole Coalescence at Redshift 0.2}},\ \href
  {https://doi.org/10.1103/PhysRevLett.118.221101} {\bibfield  {journal}
  {\bibinfo  {journal} {Phys. Rev. Lett.}\ }\textbf {\bibinfo {volume} {118}},\
  \bibinfo {pages} {221101} (\bibinfo {year} {2017}{\natexlab{a}})},\ \Eprint
  {https://arxiv.org/abs/1706.01812} {arXiv:1706.01812} \BibitemShut {NoStop}%
\bibitem [{\citenamefont {Abbott}\ \emph
  {et~al.}(2017{\natexlab{b}})\citenamefont {Abbott} \emph
  {et~al.}}]{Abbott:2017oio}%
  \BibitemOpen
  \bibfield  {author} {\bibinfo {author} {\bibfnamefont {B.~P.}\ \bibnamefont
  {Abbott}} \emph {et~al.} (\bibinfo {collaboration} {Virgo, LIGO
  Scientific}),\ }\bibinfo {title} {{GW170814: A Three-Detector Observation of
  Gravitational Waves from a Binary Black Hole Coalescence}},\ \href
  {https://doi.org/10.1103/PhysRevLett.119.141101} {\bibfield  {journal}
  {\bibinfo  {journal} {Phys. Rev. Lett.}\ }\textbf {\bibinfo {volume} {119}},\
  \bibinfo {pages} {141101} (\bibinfo {year} {2017}{\natexlab{b}})},\ \Eprint
  {https://arxiv.org/abs/1709.09660} {arXiv:1709.09660} \BibitemShut {NoStop}%
\bibitem [{\citenamefont {Abbott}\ \emph
  {et~al.}(2017{\natexlab{c}})\citenamefont {Abbott} \emph
  {et~al.}}]{TheLIGOScientific:2017qsa}%
  \BibitemOpen
  \bibfield  {author} {\bibinfo {author} {\bibfnamefont {B.~P.}\ \bibnamefont
  {Abbott}} \emph {et~al.} (\bibinfo {collaboration} {Virgo, LIGO
  Scientific}),\ }\bibinfo {title} {{GW170817: Observation of Gravitational
  Waves from a Binary Neutron Star Inspiral}},\ \href
  {https://doi.org/10.1103/PhysRevLett.119.161101} {\bibfield  {journal}
  {\bibinfo  {journal} {Phys. Rev. Lett.}\ }\textbf {\bibinfo {volume} {119}},\
  \bibinfo {pages} {161101} (\bibinfo {year} {2017}{\natexlab{c}})},\ \Eprint
  {https://arxiv.org/abs/1710.05832} {arXiv:1710.05832} \BibitemShut {NoStop}%
\bibitem [{\citenamefont {Abbott}\ \emph
  {et~al.}(2017{\natexlab{d}})\citenamefont {Abbott} \emph
  {et~al.}}]{Abbott:2017gyy}%
  \BibitemOpen
  \bibfield  {author} {\bibinfo {author} {\bibfnamefont {B.~P.}\ \bibnamefont
  {Abbott}} \emph {et~al.} (\bibinfo {collaboration} {Virgo, LIGO
  Scientific}),\ }\bibinfo {title} {{GW170608: Observation of a 19-solar-mass
  Binary Black Hole Coalescence}},\ \href
  {https://doi.org/10.3847/2041-8213/aa9f0c} {\bibfield  {journal} {\bibinfo
  {journal} {Astrophys. J.}\ }\textbf {\bibinfo {volume} {851}},\ \bibinfo
  {pages} {L35} (\bibinfo {year} {2017}{\natexlab{d}})},\ \Eprint
  {https://arxiv.org/abs/1711.05578} {arXiv:1711.05578} \BibitemShut {NoStop}%
\bibitem [{\citenamefont {Abbott}\ \emph {et~al.}(2019)\citenamefont {Abbott}
  \emph {et~al.}}]{LIGOScientific:2018mvr}%
  \BibitemOpen
  \bibfield  {author} {\bibinfo {author} {\bibfnamefont {B.}~\bibnamefont
  {Abbott}} \emph {et~al.} (\bibinfo {collaboration} {LIGO Scientific,
  Virgo}),\ }\bibinfo {title} {{GWTC-1: A Gravitational-Wave Transient Catalog
  of Compact Binary Mergers Observed by LIGO and Virgo during the First and
  Second Observing Runs}},\ \href {https://doi.org/10.1103/PhysRevX.9.031040}
  {\bibfield  {journal} {\bibinfo  {journal} {Phys. Rev. X}\ }\textbf {\bibinfo
  {volume} {9}},\ \bibinfo {pages} {031040} (\bibinfo {year} {2019})},\ \Eprint
  {https://arxiv.org/abs/1811.12907} {arXiv:1811.12907} \BibitemShut {NoStop}%
\bibitem
[{\citenamefont {Akrami}\ \emph {et~al.}(2018)\citenamefont {Akrami}
  \emph {et~al.}}]{Akrami:2018odb}%
  \BibitemOpen
  \bibfield  {author} {\bibinfo {author} {\bibfnamefont {Y.}~\bibnamefont
  {Akrami}} \emph {et~al.} (\bibinfo {collaboration} {Planck}),\ }\bibinfo
  {title} {{Planck 2018 results. X. Constraints on inflation}},
  \ \href {https://doi.org/10.1103/PhysRevD.75.123518}
  {\bibfield  {journal} {\bibinfo  {journal} {Astron. Astrophys.}\ }\textbf {\bibinfo
  {volume} {641}},\ \bibinfo {pages} {A10} (\bibinfo {year} {2020})},\
   \Eprint
  {https://arxiv.org/abs/1807.06211} {arXiv:1807.06211}\
  \BibitemShut {NoStop}%
\bibitem [{\citenamefont {Ananda}\ \emph {et~al.}(2007)\citenamefont {Ananda},
  \citenamefont {Clarkson},\ and\ \citenamefont {Wands}}]{Ananda:2006af}%
  \BibitemOpen
  \bibfield  {author} {\bibinfo {author} {\bibfnamefont {K.~N.}\ \bibnamefont
  {Ananda}}, \bibinfo {author} {\bibfnamefont {C.}~\bibnamefont {Clarkson}},\
  and\ \bibinfo {author} {\bibfnamefont {D.}~\bibnamefont {Wands}},\ }\bibinfo
  {title} {{The Cosmological gravitational wave background from primordial
  density perturbations}},\ \href {https://doi.org/10.1103/PhysRevD.75.123518}
  {\bibfield  {journal} {\bibinfo  {journal} {Phys. Rev. D}\ }\textbf {\bibinfo
  {volume} {75}},\ \bibinfo {pages} {123518} (\bibinfo {year} {2007})},\
  \Eprint {https://arxiv.org/abs/gr-qc/0612013} {arXiv:gr-qc/0612013}
  \BibitemShut {NoStop}%
\bibitem [{\citenamefont {Baumann}\ \emph {et~al.}(2007)\citenamefont
  {Baumann}, \citenamefont {Steinhardt}, \citenamefont {Takahashi},\ and\
  \citenamefont {Ichiki}}]{Baumann:2007zm}%
  \BibitemOpen
  \bibfield  {author} {\bibinfo {author} {\bibfnamefont {D.}~\bibnamefont
  {Baumann}}, \bibinfo {author} {\bibfnamefont {P.~J.}\ \bibnamefont
  {Steinhardt}}, \bibinfo {author} {\bibfnamefont {K.}~\bibnamefont
  {Takahashi}},\ and\ \bibinfo {author} {\bibfnamefont {K.}~\bibnamefont
  {Ichiki}},\ }\bibinfo {title} {{Gravitational Wave Spectrum Induced by
  Primordial Scalar Perturbations}},\ \href
  {https://doi.org/10.1103/PhysRevD.76.084019} {\bibfield  {journal} {\bibinfo
  {journal} {Phys. Rev. D}\ }\textbf {\bibinfo {volume} {76}},\ \bibinfo
  {pages} {084019} (\bibinfo {year} {2007})},\ \Eprint
  {https://arxiv.org/abs/hep-th/0703290} {arXiv:hep-th/0703290} \BibitemShut
  {NoStop}%
\bibitem [{\citenamefont {Bugaev}\ and\ \citenamefont
  {Klimai}(2010)}]{Bugaev:2009zh}%
  \BibitemOpen
  \bibfield  {author} {\bibinfo {author} {\bibfnamefont {E.}~\bibnamefont
  {Bugaev}}\ and\ \bibinfo {author} {\bibfnamefont {P.}~\bibnamefont
  {Klimai}},\ }\bibinfo {title} {{Induced gravitational wave background and
  primordial black holes}},\ \href {https://doi.org/10.1103/PhysRevD.81.023517}
  {\bibfield  {journal} {\bibinfo  {journal} {Phys. Rev. D}\ }\textbf {\bibinfo
  {volume} {81}},\ \bibinfo {pages} {023517} (\bibinfo {year} {2010})},\
  \Eprint {https://arxiv.org/abs/0908.0664} {arXiv:0908.0664} \BibitemShut
  {NoStop}%
\bibitem [{\citenamefont {Bugaev}\ and\ \citenamefont
  {Klimai}(2011)}]{Bugaev:2010bb}%
  \BibitemOpen
  \bibfield  {author} {\bibinfo {author} {\bibfnamefont {E.}~\bibnamefont
  {Bugaev}}\ and\ \bibinfo {author} {\bibfnamefont {P.}~\bibnamefont
  {Klimai}},\ }\bibinfo {title} {{Constraints on the induced gravitational wave
  background from primordial black holes}},\ \href
  {https://doi.org/10.1103/PhysRevD.83.083521} {\bibfield  {journal} {\bibinfo
  {journal} {Phys. Rev. D}\ }\textbf {\bibinfo {volume} {83}},\ \bibinfo
  {pages} {083521} (\bibinfo {year} {2011})},\ \Eprint
  {https://arxiv.org/abs/1012.4697} {arXiv:1012.4697} \BibitemShut {NoStop}%
\bibitem [{\citenamefont {Inomata}\ \emph {et~al.}(2017)\citenamefont
  {Inomata}, \citenamefont {Kawasaki}, \citenamefont {Mukaida}, \citenamefont
  {Tada},\ and\ \citenamefont {Yanagida}}]{Inomata:2016rbd}%
  \BibitemOpen
  \bibfield  {author} {\bibinfo {author} {\bibfnamefont {K.}~\bibnamefont
  {Inomata}}, \bibinfo {author} {\bibfnamefont {M.}~\bibnamefont {Kawasaki}},
  \bibinfo {author} {\bibfnamefont {K.}~\bibnamefont {Mukaida}}, \bibinfo
  {author} {\bibfnamefont {Y.}~\bibnamefont {Tada}},\ and\ \bibinfo {author}
  {\bibfnamefont {T.~T.}\ \bibnamefont {Yanagida}},\ }\bibinfo {title}
  {{Inflationary primordial black holes for the LIGO gravitational wave events
  and pulsar timing array experiments}},\ \href
  {https://doi.org/10.1103/PhysRevD.95.123510} {\bibfield  {journal} {\bibinfo
  {journal} {Phys. Rev. D}\ }\textbf {\bibinfo {volume} {95}},\ \bibinfo
  {pages} {123510} (\bibinfo {year} {2017})},\ \Eprint
  {https://arxiv.org/abs/1611.06130} {arXiv:1611.06130} \BibitemShut {NoStop}%
\bibitem [{\citenamefont {Di}\ and\ \citenamefont {Gong}(2018)}]{Gong:2017qlj}%
  \BibitemOpen
  \bibfield  {author} {\bibinfo {author} {\bibfnamefont {H.}~\bibnamefont
  {Di}}\ and\ \bibinfo {author} {\bibfnamefont {Y.}~\bibnamefont {Gong}},\
  }\bibinfo {title} {{Primordial black holes and second order gravitational
  waves from ultra-slow-roll inflation}},\ \href
  {https://doi.org/10.1088/1475-7516/2018/07/007} {J. Cosmol. Astropart. Phys.\
  \bibinfo {volume} {07}\bibfield  {year} {\bibinfo  {year} { (\textbf
  {2018})}\ }\bibfield  {pages} {\bibinfo  {pages} {007}},\ }\Eprint
  {https://arxiv.org/abs/1707.09578} {arXiv:1707.09578} \BibitemShut {NoStop}%
\bibitem [{\citenamefont {Cai}\ \emph {et~al.}(2019{\natexlab{a}})\citenamefont
  {Cai}, \citenamefont {Pi},\ and\ \citenamefont {Sasaki}}]{Cai:2018dig}%
  \BibitemOpen
  \bibfield  {author} {\bibinfo {author} {\bibfnamefont {R.-G.}\ \bibnamefont
  {Cai}}, \bibinfo {author} {\bibfnamefont {S.}~\bibnamefont {Pi}},\ and\
  \bibinfo {author} {\bibfnamefont {M.}~\bibnamefont {Sasaki}},\ }\bibinfo
  {title} {{Gravitational Waves Induced by non-Gaussian Scalar
  Perturbations}},\ \href {https://doi.org/10.1103/PhysRevLett.122.201101}
  {\bibfield  {journal} {\bibinfo  {journal} {Phys. Rev. Lett.}\ }\textbf
  {\bibinfo {volume} {122}},\ \bibinfo {pages} {201101} (\bibinfo {year}
  {2019}{\natexlab{a}})},\ \Eprint {https://arxiv.org/abs/1810.11000}
  {arXiv:1810.11000} \BibitemShut {NoStop}%
\bibitem [{\citenamefont {Kohri}\ and\ \citenamefont
  {Terada}(2018)}]{Kohri:2018awv}%
  \BibitemOpen
  \bibfield  {author} {\bibinfo {author} {\bibfnamefont {K.}~\bibnamefont
  {Kohri}}\ and\ \bibinfo {author} {\bibfnamefont {T.}~\bibnamefont {Terada}},\
  }\bibinfo {title} {{Semianalytic calculation of gravitational wave spectrum
  nonlinearly induced from primordial curvature perturbations}},\ \href
  {https://doi.org/10.1103/PhysRevD.97.123532} {\bibfield  {journal} {\bibinfo
  {journal} {Phys. Rev. D}\ }\textbf {\bibinfo {volume} {97}},\ \bibinfo
  {pages} {123532} (\bibinfo {year} {2018})},\ \Eprint
  {https://arxiv.org/abs/1804.08577} {arXiv:1804.08577} \BibitemShut {NoStop}%
\bibitem [{\citenamefont {Lu}\ \emph {et~al.}(2019)\citenamefont {Lu},
  \citenamefont {Gong}, \citenamefont {Yi},\ and\ \citenamefont
  {Zhang}}]{Lu:2019sti}%
  \BibitemOpen
  \bibfield  {author} {\bibinfo {author} {\bibfnamefont {Y.}~\bibnamefont
  {Lu}}, \bibinfo {author} {\bibfnamefont {Y.}~\bibnamefont {Gong}}, \bibinfo
  {author} {\bibfnamefont {Z.}~\bibnamefont {Yi}},\ and\ \bibinfo {author}
  {\bibfnamefont {F.}~\bibnamefont {Zhang}},\ }\bibinfo {title} {{Constraints
  on primordial curvature perturbations from primordial black hole dark matter
  and secondary gravitational waves}},\ \href
  {https://doi.org/10.1088/1475-7516/2019/12/031} {J. Cosmol. Astropart. Phys.\
  \bibinfo {volume} {12}\bibfield  {year} {\bibinfo  {year} { (\textbf
  {2019})}\ }\bibfield  {pages} {\bibinfo  {pages} {031}},\ }\Eprint
  {https://arxiv.org/abs/1907.11896} {arXiv:1907.11896} \BibitemShut {NoStop}%
\bibitem [{\citenamefont {Cai}\ \emph {et~al.}(2019{\natexlab{b}})\citenamefont
  {Cai}, \citenamefont {Pi}, \citenamefont {Wang},\ and\ \citenamefont
  {Yang}}]{Cai:2019amo}%
  \BibitemOpen
  \bibfield  {author} {\bibinfo {author} {\bibfnamefont {R.-G.}\ \bibnamefont
  {Cai}}, \bibinfo {author} {\bibfnamefont {S.}~\bibnamefont {Pi}}, \bibinfo
  {author} {\bibfnamefont {S.-J.}\ \bibnamefont {Wang}},\ and\ \bibinfo
  {author} {\bibfnamefont {X.-Y.}\ \bibnamefont {Yang}},\ }\bibinfo {title}
  {{Resonant multiple peaks in the induced gravitational waves}},\ \href
  {https://doi.org/10.1088/1475-7516/2019/05/013} {J. Cosmol. Astropart. Phys.\
  \bibinfo {volume} {05}\bibfield  {year} {\bibinfo  {year} { (\textbf
  {2019})}\ }\bibfield  {pages} {\bibinfo  {pages} {013}},\ }\Eprint
  {https://arxiv.org/abs/1901.10152} {arXiv:1901.10152} \BibitemShut {NoStop}%
\bibitem [{\citenamefont {Cai}\ \emph {et~al.}(2019{\natexlab{c}})\citenamefont
  {Cai}, \citenamefont {Pi}, \citenamefont {Wang},\ and\ \citenamefont
  {Yang}}]{Cai:2019elf}%
  \BibitemOpen
  \bibfield  {author} {\bibinfo {author} {\bibfnamefont {R.-G.}\ \bibnamefont
  {Cai}}, \bibinfo {author} {\bibfnamefont {S.}~\bibnamefont {Pi}}, \bibinfo
  {author} {\bibfnamefont {S.-J.}\ \bibnamefont {Wang}},\ and\ \bibinfo
  {author} {\bibfnamefont {X.-Y.}\ \bibnamefont {Yang}},\ }\bibinfo {title}
  {{Pulsar Timing Array Constraints on the Induced Gravitational Waves}},\
  \href {https://doi.org/10.1088/1475-7516/2019/10/059} {J. Cosmol. Astropart.
  Phys.\ \bibinfo {volume} {10}\bibfield  {year} {\bibinfo  {year} { (\textbf
  {2019})}\ }\bibfield  {pages} {\bibinfo  {pages} {059}},\ }\Eprint
  {https://arxiv.org/abs/1907.06372} {arXiv:1907.06372} \BibitemShut {NoStop}%
\bibitem [{\citenamefont {Drees}\ and\ \citenamefont
  {Xu}(2019)}]{Drees:2019xpp}%
  \BibitemOpen
  \bibfield  {author} {\bibinfo {author} {\bibfnamefont {M.}~\bibnamefont
  {Drees}}\ and\ \bibinfo {author} {\bibfnamefont {Y.}~\bibnamefont {Xu}},\
  }\bibinfo {title} {{Critical Higgs Inflation and Second Order Gravitational
  Wave Signatures}},\ \Eprint {https://arxiv.org/abs/1905.13581}
  {arXiv:1905.13581} \BibitemShut {NoStop}%
\bibitem [{\citenamefont {Inomata}\ and\ \citenamefont
  {Nakama}(2019)}]{Inomata:2018epa}%
  \BibitemOpen
  \bibfield  {author} {\bibinfo {author} {\bibfnamefont {K.}~\bibnamefont
  {Inomata}}\ and\ \bibinfo {author} {\bibfnamefont {T.}~\bibnamefont
  {Nakama}},\ }\bibinfo {title} {{Gravitational waves induced by scalar
  perturbations as probes of the small-scale primordial spectrum}},\ \href
  {https://doi.org/10.1103/PhysRevD.99.043511} {\bibfield  {journal} {\bibinfo
  {journal} {Phys. Rev. D}\ }\textbf {\bibinfo {volume} {99}},\ \bibinfo
  {pages} {043511} (\bibinfo {year} {2019})},\ \Eprint
  {https://arxiv.org/abs/1812.00674} {arXiv:1812.00674} \BibitemShut {NoStop}%
\bibitem [{\citenamefont {Inomata}\ \emph
  {et~al.}(2019{\natexlab{a}})\citenamefont {Inomata}, \citenamefont {Kohri},
  \citenamefont {Nakama},\ and\ \citenamefont {Terada}}]{Inomata:2019ivs}%
  \BibitemOpen
  \bibfield  {author} {\bibinfo {author} {\bibfnamefont {K.}~\bibnamefont
  {Inomata}}, \bibinfo {author} {\bibfnamefont {K.}~\bibnamefont {Kohri}},
  \bibinfo {author} {\bibfnamefont {T.}~\bibnamefont {Nakama}},\ and\ \bibinfo
  {author} {\bibfnamefont {T.}~\bibnamefont {Terada}},\ }\bibinfo {title}
  {{Enhancement of Gravitational Waves Induced by Scalar Perturbations due to a
  Sudden Transition from an Early Matter Era to the Radiation Era}},\ \href
  {https://doi.org/10.1103/PhysRevD.100.043532} {\bibfield  {journal} {\bibinfo
   {journal} {Phys. Rev. D}\ }\textbf {\bibinfo {volume} {100}},\ \bibinfo
  {pages} {043532} (\bibinfo {year} {2019}{\natexlab{a}})},\ \Eprint
  {https://arxiv.org/abs/1904.12879} {arXiv:1904.12879} \BibitemShut {NoStop}%
\bibitem [{\citenamefont {Inomata}\ \emph
  {et~al.}(2019{\natexlab{b}})\citenamefont {Inomata}, \citenamefont {Kohri},
  \citenamefont {Nakama},\ and\ \citenamefont {Terada}}]{Inomata:2019zqy}%
  \BibitemOpen
  \bibfield  {author} {\bibinfo {author} {\bibfnamefont {K.}~\bibnamefont
  {Inomata}}, \bibinfo {author} {\bibfnamefont {K.}~\bibnamefont {Kohri}},
  \bibinfo {author} {\bibfnamefont {T.}~\bibnamefont {Nakama}},\ and\ \bibinfo
  {author} {\bibfnamefont {T.}~\bibnamefont {Terada}},\ }\bibinfo {title}
  {{Gravitational Waves Induced by Scalar Perturbations during a Gradual
  Transition from an Early Matter Era to the Radiation Era}},\ \href
  {https://doi.org/10.1088/1475-7516/2019/10/071} {J. Cosmol. Astropart. Phys.\
  \bibinfo {volume} {10}\bibfield  {year} {\bibinfo  {year} { (\textbf
  {2019})}\ }\bibfield  {pages} {\bibinfo  {pages} {071}},\ }\Eprint
  {https://arxiv.org/abs/1904.12878} {arXiv:1904.12878} \BibitemShut {NoStop}%
\bibitem [{\citenamefont {Espinosa}\ \emph {et~al.}(2018)\citenamefont
  {Espinosa}, \citenamefont {Racco},\ and\ \citenamefont
  {Riotto}}]{Espinosa:2018eve}%
  \BibitemOpen
  \bibfield  {author} {\bibinfo {author} {\bibfnamefont {J.~R.}\ \bibnamefont
  {Espinosa}}, \bibinfo {author} {\bibfnamefont {D.}~\bibnamefont {Racco}},\
  and\ \bibinfo {author} {\bibfnamefont {A.}~\bibnamefont {Riotto}},\ }\bibinfo
  {title} {{A Cosmological Signature of the SM Higgs Instability: Gravitational
  Waves}},\ \href {https://doi.org/10.1088/1475-7516/2018/09/012} {J. Cosmol.
  Astropart. Phys.\ \bibinfo {volume} {09}\bibfield  {year} {\bibinfo  {year} {
  (\textbf {2018})}\ }\bibfield  {pages} {\bibinfo  {pages} {012}},\ }\Eprint
  {https://arxiv.org/abs/1804.07732} {arXiv:1804.07732} \BibitemShut {NoStop}%
\bibitem [{\citenamefont {Orlofsky}\ \emph {et~al.}(2017)\citenamefont
  {Orlofsky}, \citenamefont {Pierce},\ and\ \citenamefont
  {Wells}}]{Orlofsky:2016vbd}%
  \BibitemOpen
  \bibfield  {author} {\bibinfo {author} {\bibfnamefont {N.}~\bibnamefont
  {Orlofsky}}, \bibinfo {author} {\bibfnamefont {A.}~\bibnamefont {Pierce}},\
  and\ \bibinfo {author} {\bibfnamefont {J.~D.}\ \bibnamefont {Wells}},\
  }\bibinfo {title} {{Inflationary theory and pulsar timing investigations of
  primordial black holes and gravitational waves}},\ \href
  {https://doi.org/10.1103/PhysRevD.95.063518} {\bibfield  {journal} {\bibinfo
  {journal} {Phys. Rev. D}\ }\textbf {\bibinfo {volume} {95}},\ \bibinfo
  {pages} {063518} (\bibinfo {year} {2017})},\ \Eprint
  {https://arxiv.org/abs/1612.05279} {arXiv:1612.05279} \BibitemShut {NoStop}%
\bibitem [{\citenamefont {Garcia-Bellido}\ \emph {et~al.}(2017)\citenamefont
  {Garcia-Bellido}, \citenamefont {Peloso},\ and\ \citenamefont
  {Unal}}]{Garcia-Bellido:2017aan}%
  \BibitemOpen
  \bibfield  {author} {\bibinfo {author} {\bibfnamefont {J.}~\bibnamefont
  {Garcia-Bellido}}, \bibinfo {author} {\bibfnamefont {M.}~\bibnamefont
  {Peloso}},\ and\ \bibinfo {author} {\bibfnamefont {C.}~\bibnamefont {Unal}},\
  }\bibinfo {title} {{Gravitational Wave signatures of inflationary models from
  Primordial Black Hole Dark Matter}},\ \href
  {https://doi.org/10.1088/1475-7516/2017/09/013} {J. Cosmol. Astropart. Phys.\
  \bibinfo {volume} {09}\bibfield  {year} {\bibinfo  {year} { (\textbf
  {2017})}\ }\bibfield  {pages} {\bibinfo  {pages} {013}},\ }\Eprint
  {https://arxiv.org/abs/1707.02441} {arXiv:1707.02441} \BibitemShut {NoStop}%
\bibitem [{\citenamefont {Garcia-Bellido}\ and\ \citenamefont
  {Ruiz~Morales}(2017)}]{Garcia-Bellido:2017mdw}%
  \BibitemOpen
  \bibfield  {author} {\bibinfo {author} {\bibfnamefont {J.}~\bibnamefont
  {Garcia-Bellido}}\ and\ \bibinfo {author} {\bibfnamefont {E.}~\bibnamefont
  {Ruiz~Morales}},\ }\bibinfo {title} {{Primordial black holes from single
  field models of inflation}},\ \href
  {https://doi.org/10.1016/j.dark.2017.09.007} {\bibfield  {journal} {\bibinfo
  {journal} {Phys. Dark Univ.}\ }\textbf {\bibinfo {volume} {18}},\ \bibinfo
  {pages} {47} (\bibinfo {year} {2017})},\ \Eprint
  {https://arxiv.org/abs/1702.03901} {arXiv:1702.03901} \BibitemShut {NoStop}%
\bibitem [{\citenamefont {Cheng}\ \emph {et~al.}(2018)\citenamefont {Cheng},
  \citenamefont {Lee},\ and\ \citenamefont {Ng}}]{Cheng:2018yyr}%
  \BibitemOpen
  \bibfield  {author} {\bibinfo {author} {\bibfnamefont {S.-L.}\ \bibnamefont
  {Cheng}}, \bibinfo {author} {\bibfnamefont {W.}~\bibnamefont {Lee}},\ and\
  \bibinfo {author} {\bibfnamefont {K.-W.}\ \bibnamefont {Ng}},\ }\bibinfo
  {title} {{Primordial black holes and associated gravitational waves in axion
  monodromy inflation}},\ \href {https://doi.org/10.1088/1475-7516/2018/07/001}
  {J. Cosmol. Astropart. Phys.\ \bibinfo {volume} {07}\bibfield  {year}
  {\bibinfo  {year} { (\textbf {2018})}\ }\bibfield  {pages} {\bibinfo  {pages}
  {001}},\ }\Eprint {https://arxiv.org/abs/1801.09050} {arXiv:1801.09050}
  \BibitemShut {NoStop}%
\bibitem [{\citenamefont {Cai}\ \emph {et~al.}(2019{\natexlab{d}})\citenamefont
  {Cai}, \citenamefont {Chen}, \citenamefont {Tong}, \citenamefont {Wang},\
  and\ \citenamefont {Yan}}]{Cai:2019jah}%
  \BibitemOpen
  \bibfield  {author} {\bibinfo {author} {\bibfnamefont {Y.-F.}\ \bibnamefont
  {Cai}}, \bibinfo {author} {\bibfnamefont {C.}~\bibnamefont {Chen}}, \bibinfo
  {author} {\bibfnamefont {X.}~\bibnamefont {Tong}}, \bibinfo {author}
  {\bibfnamefont {D.-G.}\ \bibnamefont {Wang}},\ and\ \bibinfo {author}
  {\bibfnamefont {S.-F.}\ \bibnamefont {Yan}},\ }\bibinfo {title} {{When
  Primordial Black Holes from Sound Speed Resonance Meet a Stochastic
  Background of Gravitational Waves}},\ \href
  {https://doi.org/10.1103/PhysRevD.100.043518} {\bibfield  {journal} {\bibinfo
   {journal} {Phys. Rev. D}\ }\textbf {\bibinfo {volume} {100}},\ \bibinfo
  {pages} {043518} (\bibinfo {year} {2019}{\natexlab{d}})},\ \Eprint
  {https://arxiv.org/abs/1902.08187} {arXiv:1902.08187} \BibitemShut {NoStop}%
\bibitem [{\citenamefont {Yuan}\ \emph
  {et~al.}(2020{\natexlab{a}})\citenamefont {Yuan}, \citenamefont {Chen},\ and\
  \citenamefont {Huang}}]{Yuan:2019wwo}%
  \BibitemOpen
  \bibfield  {author} {\bibinfo {author} {\bibfnamefont {C.}~\bibnamefont
  {Yuan}}, \bibinfo {author} {\bibfnamefont {Z.-C.}\ \bibnamefont {Chen}},\
  and\ \bibinfo {author} {\bibfnamefont {Q.-G.}\ \bibnamefont {Huang}},\
  }\bibinfo {title} {{Log-dependent slope of scalar induced gravitational waves
  in the infrared regions}},\ \href
  {https://doi.org/10.1103/PhysRevD.101.043019} {\bibfield  {journal} {\bibinfo
   {journal} {Phys. Rev. D}\ }\textbf {\bibinfo {volume} {101}},\ \bibinfo
  {pages} {043019} (\bibinfo {year} {2020}{\natexlab{a}})},\ \Eprint
  {https://arxiv.org/abs/1910.09099} {arXiv:1910.09099} \BibitemShut {NoStop}%
\bibitem [{\citenamefont {De~Luca}\ \emph {et~al.}(2020)\citenamefont
  {De~Luca}, \citenamefont {Franciolini}, \citenamefont {Kehagias},\ and\
  \citenamefont {Riotto}}]{DeLuca:2019ufz}%
  \BibitemOpen
  \bibfield  {author} {\bibinfo {author} {\bibfnamefont {V.}~\bibnamefont
  {De~Luca}}, \bibinfo {author} {\bibfnamefont {G.}~\bibnamefont
  {Franciolini}}, \bibinfo {author} {\bibfnamefont {A.}~\bibnamefont
  {Kehagias}},\ and\ \bibinfo {author} {\bibfnamefont {A.}~\bibnamefont
  {Riotto}},\ }\bibinfo {title} {{On the Gauge Invariance of Cosmological
  Gravitational Waves}},\ \href {https://doi.org/10.1088/1475-7516/2020/03/014}
  {J. Cosmol. Astropart. Phys.\ \bibinfo {volume} {03}\bibfield  {year}
  {\bibinfo  {year} { (\textbf {2020})}\ }\bibfield  {pages} {\bibinfo  {pages}
  {014}},\ }\Eprint {https://arxiv.org/abs/1911.09689} {arXiv:1911.09689}
  \BibitemShut {NoStop}%
\bibitem [{\citenamefont {Fu}\ \emph {et~al.}(2020)\citenamefont {Fu},
  \citenamefont {Wu},\ and\ \citenamefont {Yu}}]{Fu:2019vqc}%
  \BibitemOpen
  \bibfield  {author} {\bibinfo {author} {\bibfnamefont {C.}~\bibnamefont
  {Fu}}, \bibinfo {author} {\bibfnamefont {P.}~\bibnamefont {Wu}},\ and\
  \bibinfo {author} {\bibfnamefont {H.}~\bibnamefont {Yu}},\ }\bibinfo {title}
  {{Scalar induced gravitational waves in inflation with gravitationally
  enhanced friction}},\ \href {https://doi.org/10.1103/PhysRevD.101.023529}
  {\bibfield  {journal} {\bibinfo  {journal} {Phys. Rev. D}\ }\textbf {\bibinfo
  {volume} {101}},\ \bibinfo {pages} {023529} (\bibinfo {year} {2020})},\
  \Eprint {https://arxiv.org/abs/1912.05927} {arXiv:1912.05927} \BibitemShut
  {NoStop}%
\bibitem [{\citenamefont {Hajkarim}\ and\ \citenamefont
  {Schaffner-Bielich}(2020)}]{Hajkarim:2019nbx}%
  \BibitemOpen
  \bibfield  {author} {\bibinfo {author} {\bibfnamefont {F.}~\bibnamefont
  {Hajkarim}}\ and\ \bibinfo {author} {\bibfnamefont {J.}~\bibnamefont
  {Schaffner-Bielich}},\ }\bibinfo {title} {{Thermal History of the Early
  Universe and Primordial Gravitational Waves from Induced Scalar
  Perturbations}},\ \href {https://doi.org/10.1103/PhysRevD.101.043522}
  {\bibfield  {journal} {\bibinfo  {journal} {Phys. Rev. D}\ }\textbf {\bibinfo
  {volume} {101}},\ \bibinfo {pages} {043522} (\bibinfo {year} {2020})},\
  \Eprint {https://arxiv.org/abs/1910.12357} {arXiv:1910.12357} \BibitemShut
  {NoStop}%
\bibitem [{\citenamefont {Cai}\ \emph {et~al.}(2019{\natexlab{e}})\citenamefont
  {Cai}, \citenamefont {Pi},\ and\ \citenamefont {Sasaki}}]{Cai:2019cdl}%
  \BibitemOpen
  \bibfield  {author} {\bibinfo {author} {\bibfnamefont {R.-G.}\ \bibnamefont
  {Cai}}, \bibinfo {author} {\bibfnamefont {S.}~\bibnamefont {Pi}},\ and\
  \bibinfo {author} {\bibfnamefont {M.}~\bibnamefont {Sasaki}},\ }\bibinfo
  {title} {{Universal infrared scaling of gravitational wave background
  spectra}},\ \Eprint {https://arxiv.org/abs/1909.13728} {arXiv:1909.13728}
  \BibitemShut {NoStop}%
\bibitem [{\citenamefont {Dom\`{e}nech}(2020)}]{Domenech:2019quo}%
  \BibitemOpen
  \bibfield  {author} {\bibinfo {author} {\bibfnamefont {G.}~\bibnamefont
  {Dom\`{e}nech}},\ }\bibinfo {title} {{Induced gravitational waves in a
  general cosmological background}},\ \href
  {https://doi.org/10.1142/S0218271820500285} {\bibfield  {journal} {\bibinfo
  {journal} {Int. J. Mod. Phys. D}\ }\textbf {\bibinfo {volume} {29}},\
  \bibinfo {pages} {2050028} (\bibinfo {year} {2020})},\ \Eprint
  {https://arxiv.org/abs/1912.05583} {arXiv:1912.05583} \BibitemShut {NoStop}%
\bibitem [{\citenamefont {Lin}\ \emph {et~al.}(2020)\citenamefont {Lin},
  \citenamefont {Gao}, \citenamefont {Gong}, \citenamefont {Lu}, \citenamefont
  {Zhang},\ and\ \citenamefont {Zhang}}]{Lin:2020goi}%
  \BibitemOpen
  \bibfield  {author} {\bibinfo {author} {\bibfnamefont {J.}~\bibnamefont
  {Lin}}, \bibinfo {author} {\bibfnamefont {Q.}~\bibnamefont {Gao}}, \bibinfo
  {author} {\bibfnamefont {Y.}~\bibnamefont {Gong}}, \bibinfo {author}
  {\bibfnamefont {Y.}~\bibnamefont {Lu}}, \bibinfo {author} {\bibfnamefont
  {C.}~\bibnamefont {Zhang}},\ and\ \bibinfo {author} {\bibfnamefont
  {F.}~\bibnamefont {Zhang}},\ }\bibinfo {title} {{Primordial black holes and
  secondary gravitational waves from $k$ and $G$ inflation}},\ \href
  {https://doi.org/10.1103/PhysRevD.101.103515} {\bibfield  {journal} {\bibinfo
   {journal} {Phys. Rev. D}\ }\textbf {\bibinfo {volume} {101}},\ \bibinfo
  {pages} {103515} (\bibinfo {year} {2020})},\ \Eprint
  {https://arxiv.org/abs/2001.05909} {arXiv:2001.05909} \BibitemShut {NoStop}%
\bibitem [{\citenamefont {Dom\`{e}nech}\ \emph {et~al.}(2020)\citenamefont
  {Dom\`{e}nech}, \citenamefont {Pi},\ and\ \citenamefont
  {Sasaki}}]{Domenech:2020kqm}%
  \BibitemOpen
  \bibfield  {author} {\bibinfo {author} {\bibfnamefont {G.}~\bibnamefont
  {Dom\`{e}nech}}, \bibinfo {author} {\bibfnamefont {S.}~\bibnamefont {Pi}},\
  and\ \bibinfo {author} {\bibfnamefont {M.}~\bibnamefont {Sasaki}},\ }\bibinfo
  {title} {{Induced gravitational waves as a probe of thermal history of the
  universe}},\ \Eprint
  {https://arxiv.org/abs/2005.12314} {arXiv:2005.12314} \BibitemShut {NoStop}%
\bibitem [{\citenamefont {Braglia}\ \emph {et~al.}(2020)\citenamefont
  {Braglia}, \citenamefont {Hazra}, \citenamefont {Finelli}, \citenamefont
  {Smoot}, \citenamefont {Sriramkumar},\ and\ \citenamefont
  {Starobinsky}}]{Braglia:2020eai}%
  \BibitemOpen
  \bibfield  {author} {\bibinfo {author} {\bibfnamefont {M.}~\bibnamefont
  {Braglia}}, \bibinfo {author} {\bibfnamefont {D.~K.}\ \bibnamefont {Hazra}},
  \bibinfo {author} {\bibfnamefont {F.}~\bibnamefont {Finelli}}, \bibinfo
  {author} {\bibfnamefont {G.~F.}\ \bibnamefont {Smoot}}, \bibinfo {author}
  {\bibfnamefont {L.}~\bibnamefont {Sriramkumar}},\ and\ \bibinfo {author}
  {\bibfnamefont {A.~A.}\ \bibnamefont {Starobinsky}},\ }\bibinfo {title}
  {{Generating PBHs and small-scale GWs in two-field models of inflation}},\ \Eprint
  {https://arxiv.org/abs/2005.02895} {arXiv:2005.02895} \BibitemShut {NoStop}%
\bibitem [{\citenamefont {Danzmann}(1997)}]{Danzmann:1997hm}%
  \BibitemOpen
  \bibfield  {author} {\bibinfo {author} {\bibfnamefont {K.}~\bibnamefont
  {Danzmann}},\ }\bibinfo {title} {{LISA: An ESA cornerstone mission for a
  gravitational wave observatory}},\ \href
  {https://doi.org/10.1088/0264-9381/14/6/002} {\bibfield  {journal} {\bibinfo
  {journal} {Class. Quant. Grav.}\ }\textbf {\bibinfo {volume} {14}},\ \bibinfo
  {pages} {1399} (\bibinfo {year} {1997})}\BibitemShut {NoStop}%
\bibitem [{\citenamefont {Audley}\ \emph {et~al.}(2017)\citenamefont {Audley}
  \emph {et~al.}}]{Audley:2017drz}%
  \BibitemOpen
  \bibfield  {author} {\bibinfo {author} {\bibfnamefont {H.}~\bibnamefont
  {Audley}} \emph {et~al.},\ }\bibinfo {title} {{Laser Interferometer Space
  Antenna}},\ \Eprint {https://arxiv.org/abs/1702.00786} {arXiv:1702.00786}
  \BibitemShut {NoStop}%
\bibitem [{\citenamefont {Luo}\ \emph {et~al.}(2016)\citenamefont {Luo} \emph
  {et~al.}}]{Luo:2015ght}%
  \BibitemOpen
  \bibfield  {author} {\bibinfo {author} {\bibfnamefont {J.}~\bibnamefont
  {Luo}} \emph {et~al.} (\bibinfo {collaboration} {TianQin}),\ }\bibinfo
  {title} {{TianQin: a space-borne gravitational wave detector}},\ \href
  {https://doi.org/10.1088/0264-9381/33/3/035010} {\bibfield  {journal}
  {\bibinfo  {journal} {Class. Quant. Grav.}\ }\textbf {\bibinfo {volume}
  {33}},\ \bibinfo {pages} {035010} (\bibinfo {year} {2016})},\ \Eprint
  {https://arxiv.org/abs/1512.02076} {arXiv:1512.02076} \BibitemShut {NoStop}%
\bibitem [{\citenamefont {Hu}\ and\ \citenamefont {Wu}(2017)}]{Hu:2017mde}%
  \BibitemOpen
  \bibfield  {author} {\bibinfo {author} {\bibfnamefont {W.-R.}\ \bibnamefont
  {Hu}}\ and\ \bibinfo {author} {\bibfnamefont {Y.-L.}\ \bibnamefont {Wu}},\
  }\bibinfo {title} {{The Taiji Program in Space for gravitational wave physics
  and the nature of gravity}},\ \href {https://doi.org/10.1093/nsr/nwx116}
  {\bibfield  {journal} {\bibinfo  {journal} {Natl. Sci. Rev.}\ }\textbf
  {\bibinfo {volume} {4}},\ \bibinfo {pages} {685} (\bibinfo {year}
  {2017})}\BibitemShut {NoStop}%
\bibitem [{\citenamefont {Kramer}\ and\ \citenamefont
  {Champion}(2013)}]{Kramer:2013kea}%
  \BibitemOpen
  \bibfield  {author} {\bibinfo {author} {\bibfnamefont {M.}~\bibnamefont
  {Kramer}}\ and\ \bibinfo {author} {\bibfnamefont {D.~J.}\ \bibnamefont
  {Champion}},\ }\bibinfo {title} {{The European Pulsar Timing Array and the
  Large European Array for Pulsars}},\ \href
  {https://doi.org/10.1088/0264-9381/30/22/224009} {\bibfield  {journal}
  {\bibinfo  {journal} {Class. Quant. Grav.}\ }\textbf {\bibinfo {volume}
  {30}},\ \bibinfo {pages} {224009} (\bibinfo {year} {2013})}\BibitemShut
  {NoStop}%
\bibitem [{\citenamefont {Hobbs}\ \emph {et~al.}(2010)\citenamefont {Hobbs}
  \emph {et~al.}}]{Hobbs:2009yy}%
  \BibitemOpen
  \bibfield  {author} {\bibinfo {author} {\bibfnamefont {G.}~\bibnamefont
  {Hobbs}} \emph {et~al.},\ }\bibfield  {booktitle} {\emph {\bibinfo
  {booktitle} {{Gravitational waves. Proceedings, 8th Edoardo Amaldi
  Conference, Amaldi 8, New York, USA, June 22-26, 2009}}},\ }\bibinfo {title}
  {{The international pulsar timing array project: using pulsars as a
  gravitational wave detector}},\ \href
  {https://doi.org/10.1088/0264-9381/27/8/084013} {\bibfield  {journal}
  {\bibinfo  {journal} {Class. Quant. Grav.}\ }\textbf {\bibinfo {volume}
  {27}},\ \bibinfo {pages} {084013} (\bibinfo {year} {2010})},\ \Eprint
  {https://arxiv.org/abs/0911.5206} {arXiv:0911.5206} \BibitemShut {NoStop}%
\bibitem [{\citenamefont {McLaughlin}(2013)}]{McLaughlin:2013ira}%
  \BibitemOpen
  \bibfield  {author} {\bibinfo {author} {\bibfnamefont {M.~A.}\ \bibnamefont
  {McLaughlin}},\ }\bibinfo {title} {{The North American Nanohertz Observatory
  for Gravitational Waves}},\ \href
  {https://doi.org/10.1088/0264-9381/30/22/224008} {\bibfield  {journal}
  {\bibinfo  {journal} {Class. Quant. Grav.}\ }\textbf {\bibinfo {volume}
  {30}},\ \bibinfo {pages} {224008} (\bibinfo {year} {2013})},\ \Eprint
  {https://arxiv.org/abs/1310.0758} {arXiv:1310.0758} \BibitemShut {NoStop}%
\bibitem [{\citenamefont {Hobbs}(2013)}]{Hobbs:2013aka}%
  \BibitemOpen
  \bibfield  {author} {\bibinfo {author} {\bibfnamefont {G.}~\bibnamefont
  {Hobbs}},\ }\bibinfo {title} {{The Parkes Pulsar Timing Array}},\ \href
  {https://doi.org/10.1088/0264-9381/30/22/224007} {\bibfield  {journal}
  {\bibinfo  {journal} {Class. Quant. Grav.}\ }\textbf {\bibinfo {volume}
  {30}},\ \bibinfo {pages} {224007} (\bibinfo {year} {2013})},\ \Eprint
  {https://arxiv.org/abs/1307.2629} {arXiv:1307.2629} \BibitemShut {NoStop}%
\bibitem [{\citenamefont {Moore}\ \emph {et~al.}(2015)\citenamefont {Moore},
  \citenamefont {Cole},\ and\ \citenamefont {Berry}}]{Moore:2014lga}%
  \BibitemOpen
  \bibfield  {author} {\bibinfo {author} {\bibfnamefont {C.~J.}\ \bibnamefont
  {Moore}}, \bibinfo {author} {\bibfnamefont {R.~H.}\ \bibnamefont {Cole}},\
  and\ \bibinfo {author} {\bibfnamefont {C.~P.~L.}\ \bibnamefont {Berry}},\
  }\bibinfo {title} {{Gravitational-wave sensitivity curves}},\ \href
  {https://doi.org/10.1088/0264-9381/32/1/015014} {\bibfield  {journal}
  {\bibinfo  {journal} {Class. Quant. Grav.}\ }\textbf {\bibinfo {volume}
  {32}},\ \bibinfo {pages} {015014} (\bibinfo {year} {2015})},\ \Eprint
  {https://arxiv.org/abs/1408.0740} {arXiv:1408.0740} \BibitemShut {NoStop}%
\bibitem [{\citenamefont {Gong}(2019)}]{Gong:2019mui}%
  \BibitemOpen
  \bibfield  {author} {\bibinfo {author} {\bibfnamefont {J.-O.}\ \bibnamefont
  {Gong}},\ }\bibinfo {title} {{Analytic integral solutions for induced
  gravitational waves}},\ \Eprint {https://arxiv.org/abs/1909.12708}
  {arXiv:1909.12708} \BibitemShut {NoStop}%
\bibitem [{\citenamefont {Matarrese}\ \emph {et~al.}(1998)\citenamefont
  {Matarrese}, \citenamefont {Mollerach},\ and\ \citenamefont
  {Bruni}}]{Matarrese:1997ay}%
  \BibitemOpen
  \bibfield  {author} {\bibinfo {author} {\bibfnamefont {S.}~\bibnamefont
  {Matarrese}}, \bibinfo {author} {\bibfnamefont {S.}~\bibnamefont
  {Mollerach}},\ and\ \bibinfo {author} {\bibfnamefont {M.}~\bibnamefont
  {Bruni}},\ }\bibinfo {title} {{Second order perturbations of the Einstein-de
  Sitter universe}},\ \href {https://doi.org/10.1103/PhysRevD.58.043504}
  {\bibfield  {journal} {\bibinfo  {journal} {Phys. Rev. D}\ }\textbf {\bibinfo
  {volume} {58}},\ \bibinfo {pages} {043504} (\bibinfo {year} {1998})},\
  \Eprint {https://arxiv.org/abs/astro-ph/9707278} {arXiv:astro-ph/9707278}
  \BibitemShut {NoStop}%
\bibitem [{\citenamefont {Bruni}\ \emph {et~al.}(1997)\citenamefont {Bruni},
  \citenamefont {Matarrese}, \citenamefont {Mollerach},\ and\ \citenamefont
  {Sonego}}]{Bruni:1996im}%
  \BibitemOpen
  \bibfield  {author} {\bibinfo {author} {\bibfnamefont {M.}~\bibnamefont
  {Bruni}}, \bibinfo {author} {\bibfnamefont {S.}~\bibnamefont {Matarrese}},
  \bibinfo {author} {\bibfnamefont {S.}~\bibnamefont {Mollerach}},\ and\
  \bibinfo {author} {\bibfnamefont {S.}~\bibnamefont {Sonego}},\ }\bibinfo
  {title} {{Perturbations of space-time: Gauge transformations and gauge
  invariance at second order and beyond}},\ \href
  {https://doi.org/10.1088/0264-9381/14/9/014} {\bibfield  {journal} {\bibinfo
  {journal} {Class. Quant. Grav.}\ }\textbf {\bibinfo {volume} {14}},\ \bibinfo
  {pages} {2585} (\bibinfo {year} {1997})},\ \Eprint
  {https://arxiv.org/abs/gr-qc/9609040} {arXiv:gr-qc/9609040} \BibitemShut
  {NoStop}%
\bibitem [{\citenamefont {Malik}\ and\ \citenamefont
  {Wands}(2009)}]{Malik:2008im}%
  \BibitemOpen
  \bibfield  {author} {\bibinfo {author} {\bibfnamefont {K.~A.}\ \bibnamefont
  {Malik}}\ and\ \bibinfo {author} {\bibfnamefont {D.}~\bibnamefont {Wands}},\
  }\bibinfo {title} {{Cosmological perturbations}},\ \href
  {https://doi.org/10.1016/j.physrep.2009.03.001} {\bibfield  {journal}
  {\bibinfo  {journal} {Phys. Rept.}\ }\textbf {\bibinfo {volume} {475}},\
  \bibinfo {pages} {1} (\bibinfo {year} {2009})},\ \Eprint
  {https://arxiv.org/abs/0809.4944} {arXiv:0809.4944} \BibitemShut {NoStop}%
\bibitem [{\citenamefont {Hwang}\ \emph {et~al.}(2017)\citenamefont {Hwang},
  \citenamefont {Jeong},\ and\ \citenamefont {Noh}}]{Hwang:2017oxa}%
  \BibitemOpen
  \bibfield  {author} {\bibinfo {author} {\bibfnamefont {J.-C.}\ \bibnamefont
  {Hwang}}, \bibinfo {author} {\bibfnamefont {D.}~\bibnamefont {Jeong}},\ and\
  \bibinfo {author} {\bibfnamefont {H.}~\bibnamefont {Noh}},\ }\bibinfo {title}
  {{Gauge dependence of gravitational waves generated from scalar
  perturbations}},\ \href {https://doi.org/10.3847/1538-4357/aa74be} {\bibfield
   {journal} {\bibinfo  {journal} {Astrophys. J.}\ }\textbf {\bibinfo {volume}
  {842}},\ \bibinfo {pages} {46} (\bibinfo {year} {2017})},\ \Eprint
  {https://arxiv.org/abs/1704.03500} {arXiv:1704.03500} \BibitemShut {NoStop}%
\bibitem [{\citenamefont {Dom\`{e}nech}\ and\ \citenamefont
  {Sasaki}(2018)}]{Domenech:2017ems}%
  \BibitemOpen
  \bibfield  {author} {\bibinfo {author} {\bibfnamefont {G.}~\bibnamefont
  {Dom\`{e}nech}}\ and\ \bibinfo {author} {\bibfnamefont {M.}~\bibnamefont
  {Sasaki}},\ }\bibinfo {title} {{Hamiltonian approach to second order gauge
  invariant cosmological perturbations}},\ \href
  {https://doi.org/10.1103/PhysRevD.97.023521} {\bibfield  {journal} {\bibinfo
  {journal} {Phys. Rev. D}\ }\textbf {\bibinfo {volume} {97}},\ \bibinfo
  {pages} {023521} (\bibinfo {year} {2018})},\ \Eprint
  {https://arxiv.org/abs/1709.09804} {arXiv:1709.09804} \BibitemShut {NoStop}%
\bibitem [{\citenamefont {Yuan}\ \emph
  {et~al.}(2020{\natexlab{b}})\citenamefont {Yuan}, \citenamefont {Chen},\ and\
  \citenamefont {Huang}}]{Yuan:2019fwv}%
  \BibitemOpen
  \bibfield  {author} {\bibinfo {author} {\bibfnamefont {C.}~\bibnamefont
  {Yuan}}, \bibinfo {author} {\bibfnamefont {Z.-C.}\ \bibnamefont {Chen}},\
  and\ \bibinfo {author} {\bibfnamefont {Q.-G.}\ \bibnamefont {Huang}},\
  }\bibinfo {title} {{Scalar induced gravitational waves in different
  gauges}},\ \href {https://doi.org/10.1103/PhysRevD.101.063018} {\bibfield
  {journal} {\bibinfo  {journal} {Phys. Rev. D}\ }\textbf {\bibinfo {volume}
  {101}},\ \bibinfo {pages} {063018} (\bibinfo {year} {2020}{\natexlab{b}})},\
  \Eprint {https://arxiv.org/abs/1912.00885} {arXiv:1912.00885} \BibitemShut
  {NoStop}%
\bibitem [{\citenamefont {Tomikawa}\ and\ \citenamefont
  {Kobayashi}(2020)}]{Tomikawa:2019tvi}%
  \BibitemOpen
  \bibfield  {author} {\bibinfo {author} {\bibfnamefont {K.}~\bibnamefont
  {Tomikawa}}\ and\ \bibinfo {author} {\bibfnamefont {T.}~\bibnamefont
  {Kobayashi}},\ }\bibinfo {title} {{On the gauge dependence of gravitational
  waves generated at second order from scalar perturbations}},\ \href
  {https://doi.org/10.1103/PhysRevD.101.083529} {\bibfield  {journal} {\bibinfo
   {journal} {Phys. Rev. D}\ }\textbf {\bibinfo {volume} {101}},\ \bibinfo
  {pages} {083529} (\bibinfo {year} {2020})},\ \Eprint
  {https://arxiv.org/abs/1910.01880} {arXiv:1910.01880} \BibitemShut {NoStop}%
\bibitem [{\citenamefont {Inomata}\ and\ \citenamefont
  {Terada}(2020)}]{Inomata:2019yww}%
  \BibitemOpen
  \bibfield  {author} {\bibinfo {author} {\bibfnamefont {K.}~\bibnamefont
  {Inomata}}\ and\ \bibinfo {author} {\bibfnamefont {T.}~\bibnamefont
  {Terada}},\ }\bibinfo {title} {{Gauge Independence of Induced Gravitational
  Waves}},\ \href {https://doi.org/10.1103/PhysRevD.101.023523} {\bibfield
  {journal} {\bibinfo  {journal} {Phys. Rev. D}\ }\textbf {\bibinfo {volume}
  {101}},\ \bibinfo {pages} {023523} (\bibinfo {year} {2020})},\ \Eprint
  {https://arxiv.org/abs/1912.00785} {arXiv:1912.00785} \BibitemShut {NoStop}%
\bibitem [{\citenamefont {Martin-Garcia}(2002)}]{xact}%
  \BibitemOpen
  \bibfield  {author} {\bibinfo {author} {\bibfnamefont {J. M.}~\bibnamefont
  {Mart\'{i}n-Garc\'{i}a}},\ }\bibinfo
  {title} {{xAct: efficient tensor computer algebra for mathematica}},\
  \href {https://www.xact.es} {\bibfield  {journal}
  {\bibinfo  {journal} {www.xact.es}}} \BibitemShut {NoStop}%
\bibitem [{\citenamefont {Pitrou}\ \emph {et~al.}(2013)\citenamefont {Pitrou},
  \citenamefont {Roy},\ and\ \citenamefont {Umeh}}]{Pitrou:2013hga}%
  \BibitemOpen
  \bibfield  {author} {\bibinfo {author} {\bibfnamefont {C.}~\bibnamefont
  {Pitrou}}, \bibinfo {author} {\bibfnamefont {X.}~\bibnamefont {Roy}},\ and\
  \bibinfo {author} {\bibfnamefont {O.}~\bibnamefont {Umeh}},\ }\bibinfo
  {title} {{xPand: An algorithm for perturbing homogeneous cosmologies}},\
  \href {https://doi.org/10.1088/0264-9381/30/16/165002} {\bibfield  {journal}
  {\bibinfo  {journal} {Class. Quant. Grav.}\ }\textbf {\bibinfo {volume}
  {30}},\ \bibinfo {pages} {165002} (\bibinfo {year} {2013})},\ \Eprint
  {https://arxiv.org/abs/1302.6174} {arXiv:1302.6174} \BibitemShut {NoStop}%
\bibitem [{\citenamefont {Press}\ and\ \citenamefont
  {Vishniac}(1980)}]{Press:1980is}%
  \BibitemOpen
  \bibfield  {author} {\bibinfo {author} {\bibfnamefont {W.~H.}\ \bibnamefont
  {Press}}\ and\ \bibinfo {author} {\bibfnamefont {E.~T.}\ \bibnamefont
  {Vishniac}},\ }\bibinfo {title} {{Tenacious myths about cosmological
  perturbations larger than the horizon size}},\ \href
  {https://doi.org/10.1086/158083} {\bibfield  {journal} {\bibinfo  {journal}
  {Astrophys. J.}\ }\textbf {\bibinfo {volume} {239}},\ \bibinfo {pages} {1}
  (\bibinfo {year} {1980})}\BibitemShut {NoStop}%
\bibitem [{\citenamefont {Bucher}\ \emph {et~al.}(2000)\citenamefont {Bucher},
  \citenamefont {Moodley},\ and\ \citenamefont {Turok}}]{Bucher:1999re}%
  \BibitemOpen
  \bibfield  {author} {\bibinfo {author} {\bibfnamefont {M.}~\bibnamefont
  {Bucher}}, \bibinfo {author} {\bibfnamefont {K.}~\bibnamefont {Moodley}},\
  and\ \bibinfo {author} {\bibfnamefont {N.}~\bibnamefont {Turok}},\ }\bibinfo
  {title} {{The General primordial cosmic perturbation}},\ \href
  {https://doi.org/10.1103/PhysRevD.62.083508} {\bibfield  {journal} {\bibinfo
  {journal} {Phys. Rev. D}\ }\textbf {\bibinfo {volume} {62}},\ \bibinfo
  {pages} {083508} (\bibinfo {year} {2000})},\ \Eprint
  {https://arxiv.org/abs/astro-ph/9904231} {arXiv:astro-ph/9904231}
  \BibitemShut {NoStop}%
\bibitem [{\citenamefont {Bednarz}(1985)}]{Bednarz:1984dn}%
  \BibitemOpen
  \bibfield  {author} {\bibinfo {author} {\bibfnamefont {B.}~\bibnamefont
  {Bednarz}},\ }\bibinfo {title} {{On Difficulties in Synchronous Gauge Density
  Fluctuations}},\ \href {https://doi.org/10.1103/PhysRevD.31.2674} {\bibfield
  {journal} {\bibinfo  {journal} {Phys. Rev. D}\ }\textbf {\bibinfo {volume}
  {31}},\ \bibinfo {pages} {2674} (\bibinfo {year} {1985})}\BibitemShut
  {NoStop}%
\bibitem [{\citenamefont {Ma}\ and\ \citenamefont
  {Bertschinger}(1995)}]{Ma:1995ey}%
  \BibitemOpen
  \bibfield  {author} {\bibinfo {author} {\bibfnamefont {C.-P.}\ \bibnamefont
  {Ma}}\ and\ \bibinfo {author} {\bibfnamefont {E.}~\bibnamefont
  {Bertschinger}},\ }\bibinfo {title} {{Cosmological perturbation theory in the
  synchronous and conformal Newtonian gauges}},\ \href
  {https://doi.org/10.1086/176550} {\bibfield  {journal} {\bibinfo  {journal}
  {Astrophys. J.}\ }\textbf {\bibinfo {volume} {455}},\ \bibinfo {pages} {7}
  (\bibinfo {year} {1995})},\ \Eprint {https://arxiv.org/abs/astro-ph/9506072}
  {arXiv:astro-ph/9506072} \BibitemShut {NoStop}%
\bibitem [{\citenamefont {Bardeen}(1980)}]{Bardeen:1980kt}%
  \BibitemOpen
  \bibfield  {author} {\bibinfo {author} {\bibfnamefont {J.~M.}\ \bibnamefont
  {Bardeen}},\ }\bibinfo {title} {{Gauge Invariant Cosmological
  Perturbations}},\ \href {https://doi.org/10.1103/PhysRevD.22.1882} {\bibfield
   {journal} {\bibinfo  {journal} {Phys. Rev. D}\ }\textbf {\bibinfo {volume}
  {22}},\ \bibinfo {pages} {1882} (\bibinfo {year} {1980})}\BibitemShut
  {NoStop}%
\end{thebibliography}
\end{document}